\documentclass[paper]{JHEP3} 

\usepackage{amsfonts}
\usepackage{amsmath}
\usepackage{subfigure}
\usepackage{epsfig}
\usepackage{latexsym}
\usepackage{graphicx}
\usepackage{amssymb}
\numberwithin{equation}{section}

\def\be{\begin{equation}}
\def\ee{\end{equation}}
\def\bea{\begin{eqnarray}}
\def\eea{\end{eqnarray}}
\def\bequ{\begin{equation}}
\def\la{\langle}
\def\ra{\rangle}
\def\eequ{\end{equation}}

\renewcommand{\thefootnote}{\fnsymbol{footnote}}

\def\Z{{\mathbb{Z}}} % \Z=\mathbb{Z}

\newcommand{\eq} {equation}
\newcommand{\eqa} {eqnarray}
\newcommand{\NN} {\mbox {$\nonumber$}}

\def\dag{\dagger}

\def\sitarel#1#2{\mathrel{\mathop{\kern0pt #1}\limits_{#2}}}

\title{
Prescription for choosing an interpolating function }
\date{}
\author{ \large
Tomohisa Takimi
  \vspace*{0.5cm} \\
 Harish-Chandra Research Institute,\\ 
Chhatnag Road, Jhusi, Allahabad 211019, India\\
\email{tomo.takimi@gmail.com}
}

\preprint{HRI/ST/1413}

\abstract{
Interpolating functional method
is a powerful tool for studying the behavior of a quantity 
in the intermediate region of the parameter space of interest
by using its perturbative 
expansions at both ends.
Recently several interpolating functional methods have been proposed,
in addition to the well-known Pad\'e approximant,
namely the ``Fractional Power of Polynomial" (FPP) 
and the ``Fractional Power of Rational functions" (FPR) methods.
Since combinations of these methods also give interpolating functions,
we may end up with multitudes of the possible approaches.
So a criterion for choosing an appropriate interpolating function is 
very much needed.
In this paper, we propose reference quantities 
which can be used for choosing a good interpolating function.
In order to validate the prescription based on these
quantities, 
we study the degree of correlation between 
``the reference quantities'' and the ``actual degree of deviation between the 
interpolating function and the true function'' 
in examples where the true functions are known.}

\begin{document}
\setcounter{footnote}{0}
\renewcommand{\thefootnote}{\arabic{footnote}}
\allowdisplaybreaks

\section{Introduction}
\label{sec:intro}
%%%%%%%%%%%%%%%%%%%%%%%%%%%%%%%%%%%%%%%%%%%%%
%%%%%%%%%%%%%%%%%%%%%%%%%%%%%%%%%%%%%%%%%%%%%
%%%%%%%%%%%%%%%%%%%%%%%%%%%%%%%%%%%%%%%%%%%%%
In theoretical physics, perturbative expansions 
are very often used to analyze the behavior of 
physical quantities with respect to the parameter of interest.
But such perturbative expansions are insufficient 
for understanding the behavior of the physical quantities 
in the entire region of the parameter space.
Although numerical simulations are often applied for computing 
such quantities in a wide range of parameter region, 
they are not always very easy to work out.
In some situations, 
the expansions of the quantities in both ends
(small end and large end, where the small end corresponds to small value of 
parameter and the large end corresponds to large value of parameter) 
are known, 
then interpolating functional methods can be used 
to provide an interpolating function which can approximate 
behavior of the quantities over the entire region of the parameter space.
The Pad\'e approximant is a well-known example for such an interpolating 
method, which can be used to find an appropriate interpolating function.
\footnote{
There are several interesting papers 
\cite{Asnin:2007rw} and \cite{Banks:2013nga}
dealing with the Pad\'e approximant. 
Ref.~\cite{Asnin:2007rw} has applied the approximant 
to the studies on the negative 
eigenvalue of the Schwarzschild black hole, 
and Ref.~\cite{Banks:2013nga} has applied to the 
various quantities
in the $\mathcal{N}=4$ super Yang-Mills theory.
Ref.~\cite{Kleinert:2001ax}
has applied another type of interpolating function 
to the $O(N)$ non-linear sigma model. }
Recently other interpolating methods also have been proposed.
Namely, the ``Fractional Power of Polynomial'' (FPP) 
method~\cite{Sen:2013oza} 
\footnote{
In~\cite{Sen:2013oza,Pius:2013tla},
the FPP has been applied to string perturbation theories.
Similarly, \cite{Beem:2013hha,Alday:2013bha} has applied 
the interpolating scheme to the $\mathcal{N}=4$ super Yang-Mills theory.} 
and the ``Fractional Power of Rational function'' (FPR)~\cite{Honda:2014bza}.
Since combinations of these methods also give interpolating functions,
we may end up with multitudes of the possible approaches. 
This multitude of interpolating functions 
causes so-called the {\it ``landscape problem''} such that
we easily get lost which among multitudes of the interpolating functions. 
So a criterion for choosing an
appropriate interpolating function is inevitable. 
%Unfortunately such a criterion 
%has not been spelled out yet.
Proposing an efficient criterion is the aim of this paper.

%This is the aim of this paper.
%We should note that the interpolating functional methods are applied
%when there is no information of the true function in hand.
In this paper,
we will propose several quantities as the reference quantities
for selecting a good interpolating function.
Because the interpolating functions will be applied when the information 
of the actual function of physical quantities are absent,
these reference quantities should be constructed only by using 
the perturbative expansions and the interpolating functions.
%without the information on the true function. 
%We will call the quantities as 
%``quantities to be criteria''.
To check whether these quantities work as good references, 
we need to check the correlation 
between the set of these quantities 
and the ``actual deviation between the true function $F(g)$
and its interpolating function $G(g)$'', where $g$ 
is a parameter of interest.
To see the correlation, we will calculate the correlation coefficients between 
them.
Though also
\cite{Honda:2014bza} 
has suggested a criterion,  
we will argue that their criterion was insufficient. %and inappropriate.
Because as explained in Appendix~\ref{Sec:Honda-judge}, 
they did not analyze the above mentioned correlations properly.
%to check the criteria.

This paper is organized as follows:
In section \ref{sec:preliminary}, we introduce the interpolating functions, 
the landscape problem and need of criteria for selecting 
a good interpolating function.
In subsection \ref{Sec:What-criteria},
we introduce the correlation coefficients.
In section \ref{Sec:Suggestion-criteria}, 
we will suggest 
the %a set of two 
reference quantities for selecting
a good interpolating function.
In section \ref{Sec:Exam}, 
we examine the reference quantities 
by computing the correlation coefficients
in examples where the actual functions are known.
%We show that the criteria works well.
%than 
%the one in \cite{Honda:2014bza}.
Section \ref{Conclusion} is conclusion and summary.

\section{Preliminary}
\label{sec:preliminary}
%%%%%%%%%%%%%%%%%%%%%%%%%%%%%%%%%%%%%%%%%%%%%
%%%%%%%%%%%%%%%%%%%%%%%%%%%%%%%%%%%%%%%%%%%%%
%%%%%%%%%%%%%%%%%%%%%%%%%%%%%%%%%%%%%%%%%%%%%
\subsection{Interpolating functions}
\label{sec:Int-funcs}
Let us consider 
a function $F(g)$ defined in $g \in [0, \infty)$, 
which has $N_s$ order small-$g$ expansion $F_s^{(N_s )}(g)$ around $g=0$
and $N_{l}$ order large-$g$ expansion $F_l^{(N_l )}(g)$ around 
$g=\infty$. The forms of the expansions are
\begin{\eq}
F_s^{(N_s )}(g) = g^a \sum_{k=0}^{N_s} s_k g^k ,\quad
F_l^{(N_l )}(g) = g^b \sum_{k=0}^{N_l} l_k g^{-k} .
\label{eq:asymptotics}
\end{\eq}
%Takimi 7/19
%Here $a, b$ are rational numbers.
We expect that
\begin{align}
 F(g) - F_s^{(N_s )}(g)  &= {\cal O}(g^{a+N_s +1}),
\NN \\
F(g) - F_l^{(N_l )}(g) &= \mathcal{O}(g^{b-N_l -1}),
%Takimi 7/19
\label{eq:exact-weak-strong}
\end{align}
around $g = 0$ and $g = \infty$ respectively.
Based on these expansions,
we %would like to 
construct smooth interpolating functions
whose small-$g$ and large-$g$ expansions coincide
to the expansions 
\eqref{eq:asymptotics}
up to some orders.

%%%%%%%%%%%%%%%%%%%%%%%%%%%%%%%%%%%%%%%%%%%%%
%%%%%%%%%%%%%%%%%%%%%%%%%%%%%%%%%%%%%%%%%%%%%
\subsubsection{Pad\'e approximant}
%%%%%%%%%%%%%%%%%%%%%%%%%%%%%%%%%%%%%%%%%%%%%
%%%%%%%%%%%%%%%%%%%%%%%%%%%%%%%%%%%%%%%%%%%%%
%The Pad\'e approximant for $b-a \in \mathbb{Z}$ is given by
% Takimi 7/19
Here, we will introduce the 
Pad\'e approximant $\mathcal{P}_{m,n}(g)$
with $0 \le m \le N_{s}, 0 \le n \le N_{l}$,
whose small-$g$ and large-$g$ expansions coincide to 
$F^{(N_{s})}_{s}(g)$ and $F_{l}^{(N_l)}(g)$ up to 
${\cal O}(g^{a+m+1})$ and ${\cal O}(g^{b-n-1})$ respectively.
If $b-a \in \mathbb{Z}$ 
in \eqref{eq:asymptotics},
$\mathcal{P}_{m,n}(g)$ can be given by
\begin{\eq}
\mathcal{P}_{m,n}(g)
= s_0 g^a \frac{ 1 +\sum_{k=1}^p c_k g^k}{1 +\sum_{k=1}^q d_k g^k } ,
\label{eq:Pade}
\end{\eq}
where 
\begin{\eq}
p = \frac{m+n+1 +(b-a)}{2}  ,\quad q = \frac{m+n+1 -(b-a)}{2}  .
\end{\eq}
$c_k$ and $d_k$ in 
\eqref{eq:Pade} are determined 
such that
series expansions around $g=0$ and $g=\infty$ of 
\eqref{eq:Pade} become consistent with
the small-$g$ and large-$g$ 
expansions \eqref{eq:asymptotics} 
up to $\mathcal{O}(g^{a+m+1})$ and $\mathcal{O}(g^{b-n-1})$, respectively. 
This construction requires 
\begin{\eq}
\frac{m+n-1+b-a}{2} \in \mathbb{Z}, %.
\quad
b-a \in \mathbb{Z}, \quad m+n+1 \ge |b-a|. %.
\label{eq:Pade_constraint}
\end{\eq}
The Pad\'e approximant is reliable
only when there is no pole or singularity in \eqref{eq:Pade},
namely the denominator in \eqref{eq:Pade}
should not have any zero point in the region of interest.

%%%%%%%%%%%%%%%%%%%%%%%%%%%%%%%%%%%%%%%%%%%%%
%%%%%%%%%%%%%%%%%%%%%%%%%%%%%%%%%%%%%%%%%%%%%
\subsubsection{Fractional Power of Polynomial method (FPP)}
%%%%%%%%%%%%%%%%%%%%%%%%%%%%%%%%%%%%%%%%%%%%%
%%%%%%%%%%%%%%%%%%%%%%%%%%%%%%%%%%%%%%%%%%%%%
In \cite{Sen:2013oza}, another type of interpolating function, 
which we call the ``Fractional Power of Polynomial'' (FPP),
is given by
\begin{\eq}
F_{m,n}(g)
= s_0 g^a \Biggl[ 1 +\sum_{k=1}^m c_k g^k +\sum_{k=0}^n d_k g^{m+n+1 -k} \Biggr]^{\frac{b-a}{m+n+1}} .
\label{eq:FPP}
\end{\eq}
As in the Pad\'e approximant case, 
the coefficients $c_k$ and $d_k$ are 
determined by consistency 
between the Taylor expansions 
of \eqref{eq:FPP} %around $g =0, \infty$
and the expansions \eqref{eq:asymptotics}.
Unlike the Pad\'e approximant, 
the FPP does not have 
any constraints with respect to $m,n, a,b$ 
like \eqref{eq:Pade_constraint}.
We can trust the FPP only when the inside of parenthesis
in \eqref{eq:FPP} is always positive in the region 
under consideration.

%%%%%%%%%%%%%%%%%%%%%%%%%%%%%%%%%%%%%%%%%%%%%
%%%%%%%%%%%%%%%%%%%%%%%%%%%%%%%%%%%%%%%%%%%%%
\subsubsection{Fractional Power of Rational function method (FPR)}
%%%%%%%%%%%%%%%%%%%%%%%%%%%%%%%%%%%%%%%%%%%%%
%%%%%%%%%%%%%%%%%%%%%%%%%%%%%%%%%%%%%%%%%%%%%
There is also a class of interpolating functions
so-called ``Fractional Power of Rational functions'' (FPR)
proposed in \cite{Honda:2014bza}. 
With following values of $\alpha$, 
\begin{\eq}
\alpha = \left\{ \begin{matrix}
\frac{a-b}{2\ell +1}  & {\rm for} & m+n:{\rm even} \cr
\frac{a-b}{2\ell}  & {\rm for} & m+n:{\rm odd}
\end{matrix} \right. ,
\quad 
\left|\frac{a-b}{\alpha}\right| \le m+n+1,
\quad
{\rm with}\ \ell \in\mathbb{Z},
\label{eq:cond}
\end{\eq}
the FPR can be defined as 
%so called 
%``Fractional Power of Rational function method (FPR)"
\begin{\eq}
F_{m,n}^{(\alpha )} (g)
= s_0 g^a \Biggl[ \frac{ 1 +\sum_{k=1}^p c_k g^k}{1 +\sum_{k=1}^q d_k g^k } \Biggr]^\alpha ,
\label{eq:FPR}
\end{\eq}
where 
\begin{\eq}
p = \frac{1}{2} \left( m+n+1 -\frac{a-b}{\alpha} \right) ,\quad
q = \frac{1}{2} \left( m+n+1 +\frac{a-b}{\alpha} \right) .
\end{\eq}
As in the Pad\'e and the FPP cases, 
we determine $c_k$ and $d_k$ in 
\eqref{eq:FPR} 
by the 
consistency between its Taylor
expansions %of \eqref{eq:FPR} %around $g = 0,\infty$ and 
and the expansions \eqref{eq:asymptotics}. 
This approach requires
\begin{\eq}
%\frac{1}{2} \left( m+n+1 +\frac{a-b}{\alpha} \right) \in \mathbb{Z} ,
p,q \in \mathbb{Z}_{\geq 0}
\end{\eq}
which leads to the condition of $\alpha$ in \eqref{eq:cond}.

The Pad\'e approximant and the FPP can be regarded as the 
special cases of the FPR.
The FPR with %consistently taken 
$|\alpha| = 1$
becomes the standard Pad\'e approximant.
On the other hand, 
by taking the upper limit of $\left|\frac{a-b}{\alpha}\right|$
in \eqref{eq:cond}, 
namely by taking $ \alpha = \frac{b-a}{m+n+1}$, 
the FPR becomes the FPP.
When the function inside the parenthesis 
has singularities or takes negative values for fractional $\alpha$,
the FPR will not be a trustable scheme. 
%The FPR is smooth unless the rational function has poles.

%%%%%%%%%%%%%%%%%%%%%%%%%%%%%%%%%%%%%%%%%%%%%
%%%%%%%%%%%%%%%%%%%%%%%%%%%%%%%%%%%%%%%%%%%%%
\subsection{Landscape problem}
\label{sec:landscape}
%%%%%%%%%%%%%%%%%%%%%%%%%%%%%%%%%%%%%%%%%%%%%
%%%%%%%%%%%%%%%%%%%%%%%%%%%%%%%%%%%%%%%%%%%%%
%In the previous subsection,
%we show that we can construct various interpolating functions. 
We should note that
a linear combination of different interpolating functions 
gives a new interpolating function. For example, 
a linear combination of the FPP approximating a function $F(g)$, 
%and the Pad\'e approximant,
\begin{\eq}
a_{1} F_{2,2}(g) + a_{2} F_{2,3}(g) + a_3 F_{3,3}(g),
\quad a_{1}+a_{2}+ a_{3} =1, \quad a_{1,2,3} \ge 0,
\end{\eq} 
is also an interpolating function 
which matches with $F(g)$ up to ${\cal O}(g^{a+3})$ near $g =0$
and up to ${\cal O}(g^{b-3})$ near $\frac{1}{g} = 0$.
Since we can take $a_{1,2,3} \in \mathbb{R}$,
there are uncountably infinite number of interpolating functions.

The presence of huge %uncountably infinite 
number of interpolating functions naturally
causes following problem:
{\it How should we choose an interpolating function
from this multitudes.}
%We will get lost 
%which among multitudes of the interpolating functions.}
%This is really the situation of the proverb
%``too many cooks spoils the broth''.
This problem is called as the ``landscape problem''.
%named in~\cite{Honda:2014bza}. 
%\footnote{This problem has been 
%sharpened during early stage of collaborations with Masazumi Honda and 
%Ashoke Sen.
%We thank them for this point.
%} % So we are grateful to them for this point.}
To choose an %For using 
%these 
interpolating function efficiently,
we need to establish {\it a criterion} for selecting a good
interpolating function which has very small deviation from the true function.

\subsection{Correlation coefficient}
\label{Sec:What-criteria}
If the ``deviation between the interpolating function $G(g)$
and the true function $F(g)$'' 
(we denote it by $De$)
is smaller, 
$G(g)$ is regarded as a better interpolating function.
So for choosing a better interpolating function, 
we only have to see the $De$.
However when we apply the interpolating functional method,
there is no information of $De$ because 
we do not know the true function $F(g)$.
%of the absence of the 
%information of $F(g)$.
Hence, we have to find alternative quantities
for measuring the above mentioned deviations without knowing $F(g)$.
We will name such quantities as the {\it reference quantities} and 
we denote them by $Cr$.
So the problem of finding a good criterion has boiled down to finding a 
suitable set of reference quantities.
%The main issue for proposing the criteria is 
%finding out the suitable reference quantities.
If the proposed $Cr$ is reliable, then %based on it, 
we should be able to guess the actual deviation $De$ upto some extent.
In other words ,
%This means that 
{\it the reliable reference quantities $Cr$ must have 
strong correlation with $De$.}
It is a well-known fact that the degree of correlation between 
two data sets
can be computed just by 
calculating the correlation coefficients between them.
So the efficiency of the proposed reference quantities can be checked
by computing the correlation coefficients between the reference 
quantities $Cr$ and the actual deviation $De$.
%For checking the correlation, 
%we calculate the correlation coefficients between 
%$Cr$ and $De$.

%For the introduction of the correlation coefficient,
Let us briefly introduce the concept of the correlation 
coefficient,
which is a statistical notion. 
%Consider 
Assume that we have 
a set up where we not only know $Cr$ but also $De$. 
%We prepare 
Say we have 
many interpolating functions,
and calculated $(Cr,De)$. %corresponding %sets of quantities
% in each interpolating functions.
%Here we get a lot of data of $(Cr,De)$, 
Then we will call the sets of $(Cr,De)$ 
%which will be called 
as samples.
Using these samples, we can compute 
the correlation coefficients between $Cr$ and $De$ as
follows
\begin{equation}
\rho_{CrDe}\equiv 
\frac{\la (De - \la De \ra)(Cr - \la Cr \ra)\ra }{\sigma_{De}\sigma_{Cr}}, 
\end{equation}
where $\la De \ra$ and $\la Cr \ra$ are sample means
of $De$ and $Cr$ respectively, and 
$\sigma_{De}^2$ and $\sigma_{Cr}^2$ are the sample variances of 
$De$ and $Cr$ respectively.
This quantity is bounded within $-1 \le \rho_{CrDe} \le 1$.
If the correlation coefficient is very close to 1, then there is  
a very strong correlation between $Cr$ and $De$,
which implies that $De$ becomes bigger and bigger if $Cr$ is bigger and bigger. 
Generally if the correlation coefficient is 
stronger than 0.7, the correlation is called strong.
If the prescription based on  
the reference quantities is reliable, 
there should be a very strong correlation between 
$Cr$ and $De$.

In Appendix~\ref{Sec:Honda-judge}, 
we explain that the criterion proposed in 
\cite{Honda:2014bza} is insufficient.
Underlying reason for the insufficiency 
comes from the fact that %of the discussion in \cite{Honda:2014bza}
%is that 
they did not analyze the correlation between
their reference quantities $I_s + I_l$ and actual degree of deviation.

\subsubsection{Random Sampling}
Because the number of the interpolating 
functions is uncountably infinite,
it is impossible to consider all the interpolating functions.
In such cases, we randomly extract a finite number of the interpolating 
functions as samples, by employing the {\it random sampling}.

The way of random sampling in this paper is as follows: 
First we prepare several interpolating functions
$G_{1}(g), G_{2}(g), \ldots, G_{\tilde{N}}(g)$,
which are already known.
Here the number of the interpolating functions
is $\tilde{N}$.
By using these functions, %as basis,
we consider  
linear combinations 
%denoted by  
\begin{equation}
\hat{G}(g) = \sum_{i = 1}^{\tilde{N}} c_i G_{i}(g), \qquad \sum_{i=1}^{\tilde{N}} c_{i} = 1.
\label{Random-linear}
\end{equation}
Here we generate sets of the numbers $c_{i}$ 
by using the random number generator in the Mathematica.
We should note that the linear combination becomes 
an interpolating function again.
Each set of randomly generated numbers $c_{i}, (i = 1\sim \tilde{N})$
has each corresponding interpolating function through 
\eqref{Random-linear}. Hence 
through \eqref{Random-linear}, 
we can extract sets of interpolating functions randomly
by using the randomly generated numbers $c_{i}$.

In following sections, 
in each of the explicit examples where both the $Cr$ and $De$ can be known, 
we calculate the correlation coefficient five times.
We use 10 samples for the 1st calculation, 20 samples for the 2nd one, 
30 samples for the 3rd one, 50 samples for the 4th one and 100 samples for 
the 5th calculation.

For validating the calculated correlation coefficients, 
we have to take care of the 
statistical significance also.
In this paper, we will employ the $0.01$ as the significance level.
If the correlation coefficient in each of calculations
(1st, 2nd, 3rd, 4th and 5th)
exceeds 0.765, 0.561,  0.463, 0.361 and 0.256 respectively,
each correlation coefficient is regarded as statistically significant.
For the reference quantities $Cr$ to be reliable enough, 
very strong correlation coefficients are required
(at least more than 0.7). 
So the number of the samples in these calculations should be 
enough to check the reliability of $Cr$.
%% Paste
\section{Proposal of reference quantities}
\label{Sec:Suggestion-criteria}
In this section, we give explicit forms of $De$ and $Cr$.
First, we will give definitions of $De$ which are the actual degree of 
deviation between the true function $F(g)$ and the interpolating function
$\hat{G}(g)$.
In this paper, we consider following quantities as $De$:
\begin{align}
 De_{1}(\hat{G};F) = {\rm Max}\left\{\left|F(g) - \hat{G}(g)\right|;
g \ge 0
\right\},
\label{DE1}
\\
 De_{2}(\hat{G};F) = \frac{1}{\Lambda}\int^{\Lambda}_{0} dg \left|
\frac{F(g)-\hat{G}(g)}{F(g)}
\right|,
\label{DE2}
\\
 De_{o}(\hat{G};F) = \frac{1}{\Lambda}\int^{\Lambda}_{0} dg \left|
F(g)-\hat{G}(g)
\right|.
\label{DEo}
\end{align}
Here the parameter 
$\Lambda$ is a cutoff of the integration domain
to make the integration well-defined.
It is set as $\Lambda = 1000$
throughout this paper.
For each $De_{1},De_{2}$ and $De_{o}$,
we will suggest corresponding reference quantities 
$Cr_{1},Cr_{2}$ and $Cr_{o}$ respectively.
We should remember that 
%$Cr_{1}$, $Cr_{2}$ and $Cr_{o}$ 
the reference quantities
must be constructed only by using
the perturbative expansions and interpolating functions.
We should also note that 
the $Cr_{1},Cr_{2}$ and $Cr_{o}$ should have the strong correlation with 
$De_{1},De_{2}$ and $De_{o}$ respectively. 
So for 
%$Cr_{1},Cr_{2}$ and $Cr_{o}$ 
each reference quantities
to have a strong correlation
with each corresponding actual degree of deviation,
each $Cr_{1},Cr_{2}$ and $Cr_{o}$ 
should be a quantity very similar to $De_{1},De_{2}$ and $De_{o}$ respectively.
Then in this paper, we will suggest following quantities
as $Cr_{1}, Cr_{2}$ and $Cr_{o}$,
\begin{equation}
Cr_1
= {\rm Max}\left[\left\{\left|\hat{G}(g) - F^{(N_s^{\ast})}_{s}(g)\right|
; 0 \le g \le g_{s}^{\ast}\right\}\cup 
\left\{\left|\hat{G}(g) - F^{(N_l^{\ast})}_{l}
(g)\right|
;  g \ge g_{l}^{\ast}\right\}\right],
\label{CR12}
\end{equation}
\begin{equation}
Cr_2 = \int_{0}^{g_{s}^{\ast}} 
dg\left|\frac{\hat{G}(g) - F_{s}^{(N_s^{\ast})}(g)}{F_{s}^{(N_s^{\ast})}(g)
}\right|
+
\int^{\Lambda}_{g_{l}^{\ast}} 
dg\left|\frac{\hat{G}(g) - F_{l}^{(N_l^{\ast})}(g)}{F_{l}^{(N_l^{\ast})}
(g)}\right|,
\label{CR22}
\end{equation}
\begin{equation}
Cr_o = \int_{0}^{g_{s}^{\ast}} 
dg\left|\hat{G}(g) - F_{s}^{(N_s^{\ast})}(g)
\right|
+
\int^{\Lambda}_{g_{l}^{\ast}} 
dg\left|\hat{G}(g) - F_{l}^{(N_l^{\ast})}(g)\right|.
\label{CRo2}
\end{equation}
Here 
$F_{s}^{(N_{s}^{\ast})}$
and $F_{l}^{(N_{l}^{\ast})}$ are the 
optimally truncated expansions of 
original expansions 
$F_{s}^{(N_{s})}$
and $F_{l}^{(N_{l})}$ %in \eqref{eq:asymptotics} 
respectively.
Here $N_{s}^{\ast} \le N_{s}$ and 
$N_{l}^{\ast} \le N_{l}$. 
%Optimally truncated series 
%is regarded as the closest expansion series to the true function,
%obtained by the truncation of the original expansion at the suitable order.
The detailed explanations on the orders $N_{s}^{\ast}, N_{l}^{\ast}$ 
and the optimal truncation are put in Appendix~\ref{Sec:Rel-opt}.
The domain
\begin{equation}
0 \le g  \le g_{s}^{\ast}, \quad 
g_{l}^{\ast} \le g, \quad 
\end{equation}
is called as the reliable domain. 
Inside the domain $0 \le g  \le g_{s}^{\ast}$, 
$F_{s}^{(N_s^{\ast})}(g)$ is sufficiently close to the true function,
while $F_{l}^{(N_l^{\ast})}(g)$ is sufficiently close to the true function 
in the domain $g_{l}^{\ast} \le g $.
%the truncated expansions $F_{s}^{(N_s^{\ast})}(g),F_{l}^{(N_l^{\ast})}(g)$ 
%are sufficiently close to the true function $F(g)$.
The detailed explanations of the reliable domain are also put in 
Appendix~\ref{Sec:Rel-opt}.

\section{Examination of reference quantities 
by the correlation functions}
\label{Sec:Exam}
In this section, we will check the reliability of 
reference quantities
%$Cr_{1}$
%and $Cr_{2}$ 
given by 
\eqref{CR12}, \eqref{CR22} and \eqref{CRo2}
in the explicit examples where 
both $Cr_{1,2,o}$ and $De_{1,2,o}$ can be computed.
For checking the reliability, we calculate the correlation coefficients
between $Cr_{j}$ and $De_{j}$ for $j \in \{1,2,o\}$.
%, and the coefficients 
%between $Cr_{2}$ and $De_{2}$.
In first three subsections of this section, 
we use following three kinds of true functions $F(g)$
as explicit examples at which we calculate the correlation 
coefficients:
\begin{enumerate}
 \item Functions where both the small-$g$ and large-$g$
expansions are convergent.
 \item Functions where one of the small-$g$ or large-$g$
expansions is convergent while the other is asymptotic.
 \item Functions where both the expansions are asymptotic.
\end{enumerate}
After these, we also discuss in the following true functions,
\begin{enumerate}
 \item[4.] 
Functions having sharp peak outside the 
reliable domain.
\end{enumerate}

\subsection{Functions with convergent expansions in both ends.}
\subsubsection{$(1-\frac{g}{5}+g^2)^\frac{1}{2}$}
\label{Sec:B-conv-1}
Let us discuss by using the following function
\begin{equation}
F(g) = \left(1-\frac{g}{5}+g^2 \right)^\frac{1}{2}  
\end{equation}
as the true function.
We can easily see that its small-$g$
and large-$g$ expansions are convergent. 
\footnote{By noting that
\begin{equation}
F(g) = \left(1+(-\frac{g}{5}+g^2) \right)^\frac{1}{2} 
= g \left(1+(-\frac{1}{5g}+\frac{1}{g^2}) \right)^\frac{1}{2}, 
\end{equation}
$F(g)$ can be rewritten as
\begin{align}
F(g) =& 1 + \frac{1}{2}
(-\frac{1}{5}g + g^2)
+ \sum_{n=2}^{\infty} 
%\frac{\Gamma(\frac{3}{2})}{\Gamma(n+1)
%\Gamma(\frac{3}{2}-n)} 
\frac{
(-1)^{n-1}(2n-3)!!
}{2^n n!}
\left(-\frac{1}{5}g + g^2\right)^n,
\label{F11-W-ex}
\\
F(g) =& g
 + \frac{g}{2}
\left(-\frac{1}{5g} + \frac{1}{g^2}\right)
+ g\sum_{n=2}^{\infty} 
\frac{
(-1)^{n-1}(2n-3)!!
}{2^n n!}
\left(-\frac{1}{5g} + \frac{1}{g^2}\right)^n,
\label{F11-S-ex}
\end{align}
around $g = 0$ and $g=\infty$ respectively.
From these series, the convergent radius $\tilde{g}_{s}, \tilde{g}_{l}$
are obtained by solving
\begin{equation}
\tilde{g}_{s}^2 
-\frac{1}{5}\tilde{g}_{s}
-1 = 0, \qquad 
\tilde{g}_{l}^{-2} 
-\frac{1}{5}\tilde{g}_{l}^{-1}
-1 = 0.
\end{equation}
}
Convergent radius for the small-$g$ expansion is 
$\tilde{g}_{s} \sim 1.10499$, 
and the one for the large-$g$ expansion is obtained as
$\tilde{g}_{l} \sim 0.904988$. 

Because we do not know $F(g)$ when we apply the 
interpolating functional method, 
to consider the reference quantities,
we should 
%prepare of the small-$g$ and large-$g$
%expansions only up-to finite order.
make an 
assumption that we know only its expansions
upto finite order in both ends.
%\paragraph{Starting circumstance for the discussion}
Suppose that 
we know the small-$g$ and large-$g$ expansions only up to 100-th 
order, 
%Suppose that 
where the small-$g$ expansion %up-to 100-th order
$F_{s}^{(100)}(g)$ %as well as the 
and 
large-$g$ expansion $F_{l}^{(100)}(g)$ %up-to 100-th order are 
%up-to 100-th order 
are given by
\begin{equation}
F_{s}^{(100)}(g) = \sum_{n=0}^{100} f_{s}^{(n)} g^{n},
%\label{F11-W-100}
%\\
\qquad
F_{l}^{(100)}(g) = \sum_{n=0}^{100} f_{l}^{(n)} g^{-n+1}.
\label{F11-S-100}
\end{equation}
%We consider $Cr_{1}, Cr_{2}$
Because the reference quantities should be constructed by %based only on 
the perturbative expansions and the interpolating functions only,
the values $g_{s}^{\ast}, g_{l}^{\ast},N_{s}^{\ast}$ and $N_{l}^{\ast}$
should be determined based on the expansions \eqref{F11-S-100} only.
(Also interpolating functions are made by the expansions only.)
In case that large order expansions are known, 
we can employ the fitting method 
to obtain these values \cite{Honda:2014bza}.
See also Appendix~\ref{Sec:Rel-opt}.

\begin{figure}[t]
\begin{center}
\includegraphics[width=7.4cm]{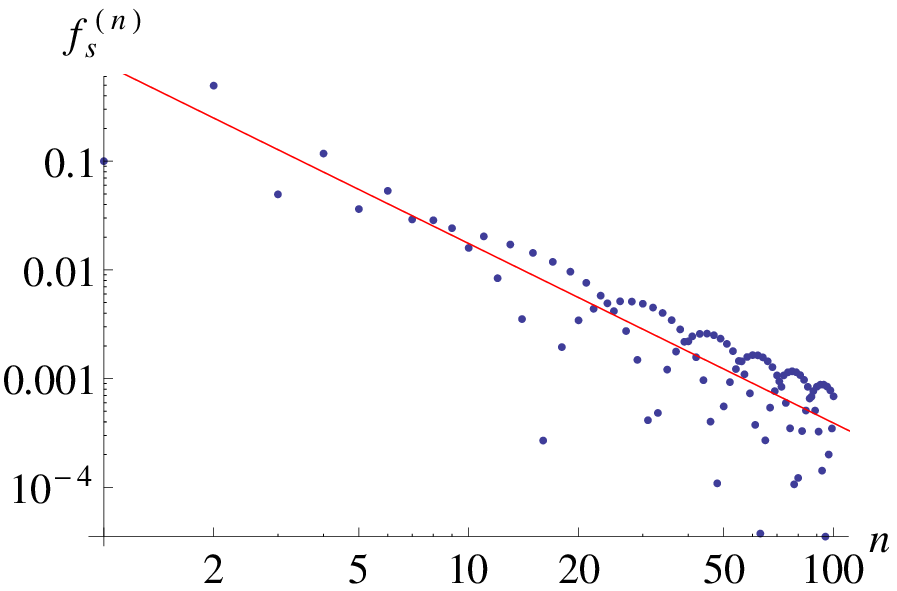}
\includegraphics[width=7.4cm]{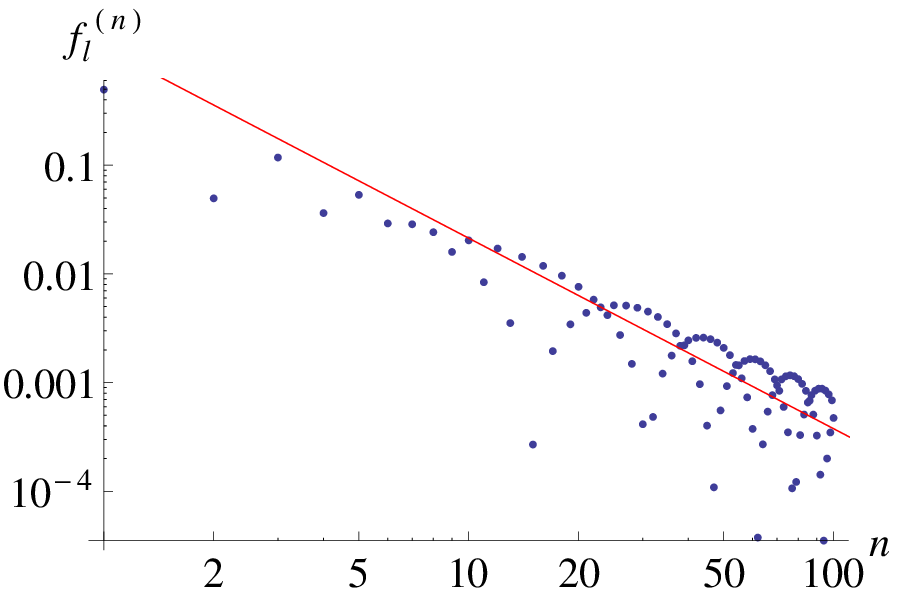}
\end{center}
\caption{
[Left] Small-$g$ expansion coefficients $|f_s^{(n)}|$ are plotted to $n$ 
in log-log scale.
[Right] Large-$g$ expansion coefficients $|f_l^{(n)}|$ are plotted to $n$ 
in log-log scale.
}
\label{fig:ex3}
\end{figure}

\begin{figure}[t]
\begin{center}
\includegraphics[width=9.4cm]{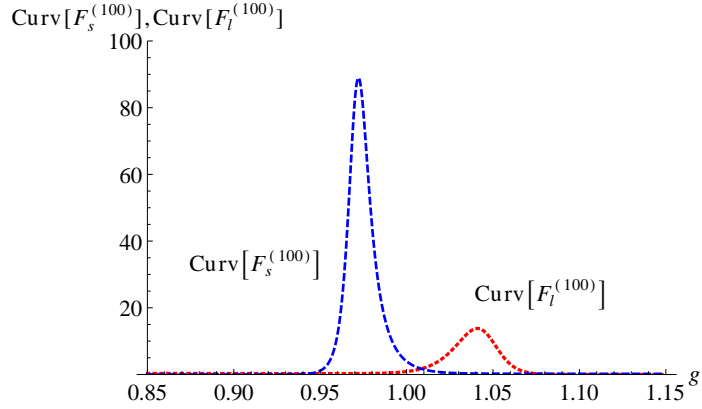}
\end{center}
\caption{
Absolute value of curvature of $F^{(100)}_{s}(g)$ and 
$F^{(100)}_{l}(g)$ with respect to $g$.
}
\label{fig:ex3-2}
\end{figure}

%\paragraph{Estimation of $g_{s}^{\ast}$ and $g_{l}^{\ast}$ 
%by fitting.}
Let us determine $g_{s}^{\ast}, g_{l}^{\ast},N_{s}^{\ast}$ and 
$N_{l}^{\ast}$ by the fitting.
\footnote{
For analysis in the Figs.~\ref{fig:ex3} and \ref{fig:ex3-2}
and Eq.~\eqref{F11-basis-list}, 
we have also utilized the analysis
made by Honda during the collaboration in the early stage. 
We thank Honda for the analysis.}
In Figure~\ref{fig:ex3}, 
we plot how the coefficients $f^{(n)}_{s}, f^{(n)}_{l}$
in \eqref{F11-S-100}
behave with respect to $n$.
We can see that these behave as $\sim n^{c}$ at large $n$ 
with some constant $c$.
So we can expect that $g_{l}^{\ast} = g_{s}^{\ast} = 1$,
and both the expansions will be convergent.
%(When we apply the interpolating functional method, 
%we have to know whether the expansion 
%is convergent or not only by the information on the finite order 
%expansions.)
In case of the convergent expansion, 
we should also take care of the blow-up point of the 
curvature.
Fig.~\ref{fig:ex3-2} plots the absolute value of the curvature
of $F^{(100)}_{s}(g)$ and $F^{(100)}_{l}(g)$
to $g$,
where the peak of the curvature of $F_{s}^{(100)}(g)$ starts from
$g = 0.95$ while the one of $F_{l}^{(100)}(g)$ starts from $g = 1.07$.
So from these observations in Figs.~\ref{fig:ex3} and \ref{fig:ex3-2},
we set 
\begin{equation}
g_{s}^{\ast} = 0.9, \qquad g_{l}^{\ast}  =1.1.
\end{equation}
Because both the small-$g$ and large-$g$ expansions are 
convergent, we will set $N_{s}^{\ast} = N_{s} = 100$
and $N_{l}^{\ast} = N_{l} = 100$.
%\paragraph{Construction of the interpolating function}
Based on the expansions
\eqref{F11-S-100},
we can construct the interpolating functions
in the ways explained in subsection \ref{sec:Int-funcs}.
We have constructed the several interpolating functions 
which are listed in \eqref{F11-basis-list}
in Appendix~\ref{Eq:F11}.

%\paragraph{Random sampling}
We consider the following linear combinations of the 
interpolating functions by using the functions 
in \eqref{F11-basis-list}% as the basis, 
\begin{align}
\hat{G}^{[r,s]}(g) 
=& c_{1}^{[r,s]}F^{(-1)}_{1,1}(g)
+c_{2}^{[r,s]}F^{(-1/3)}_{1,1}(g)
+c_{3}^{[r,s]}F^{(-1)}_{2,2}(g)
+c_{4}^{[r,s]}F^{(-1/3)}_{2,2}(g)
\NN \\
&+c_{5}^{[r,s]}F^{(-1/5)}_{2,2}(g)
+c_{6}^{[r,s]}F^{(-1)}_{3,3}(g)
+c_{7}^{[r,s]}F^{(-1/3)}_{3,3}(g)
+c_{8}^{[r,s]}F^{(-1/5)}_{3,3}(g)
\NN \\
&+c_{9}^{[r,s]}F^{(-1/7)}_{3,3}(g)
+c_{10}^{[r,s]}F^{(-1)}_{4,4}(g)
+c_{11}^{[r,s]}F^{(-1/3)}_{4,4}(g)
+c_{12}^{[r,s]}F^{(-1/5)}_{4,4}(g)
\NN \\
&+c_{13}^{[r,s]}F^{(-1/7)}_{4,4}(g)
+c_{14}^{[r,s]}F^{(-1/9)}_{4,4}(g),
\label{Eq:Int-func-conv-1}
\end{align}
where 
\begin{equation}
\sum_{i=1}^{14}c_{i}^{[r,s]} =  1, \qquad c_{i}^{[r,s]}\ge 0.
\label{constraint}
\end{equation}
Here $c_{i}^{[r,s]}$ are randomly chosen %with the above constraint
%\eqref{constraint} 
by using the random number generator in the 
Mathematica.
The superscript $[r,s]$ indicates the $r$-th
sample for the $s$-th calculation of correlation coefficients.
(We use 10 samples for 1st calculation, 
20 samples for 2nd one, 30 samples for 3rd one, 
50 samples for 4th one and 100 samples for 5th calculation.)
At least these functions match with the small-$g$ and large-$g$
expansions up to %order 1, 
\begin{equation}
F^{(100)}_{s}(g) - \hat{G}^{[r,s]}(g) = {\cal O}(g^2), \qquad
F^{(100)}_{l}(g) - \hat{G}^{[r,s]}(g) = {\cal O}(1/g).
\end{equation}
%The list of the coefficient is put in {\it where should I put ??
%in the draft, or other pdf file ?}

\begin{figure}[tbp]
\begin{center}
\includegraphics[width=7.0cm]{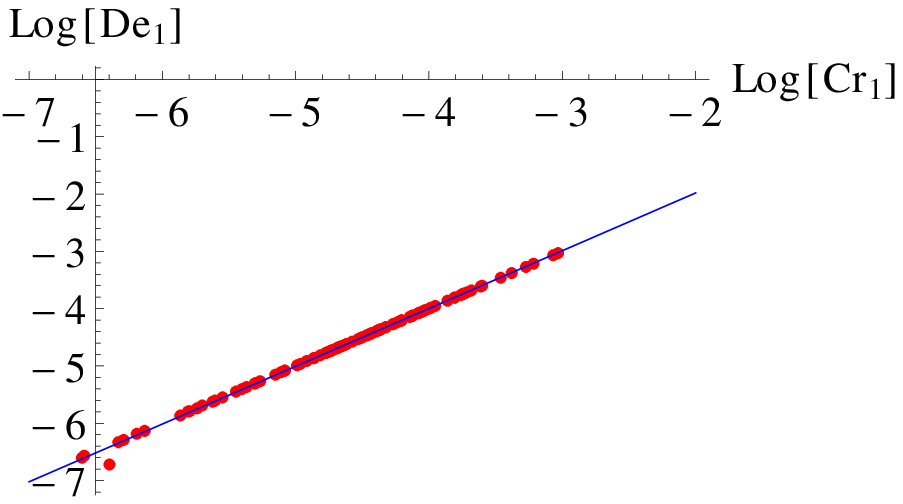}
\includegraphics[width=7.0cm]{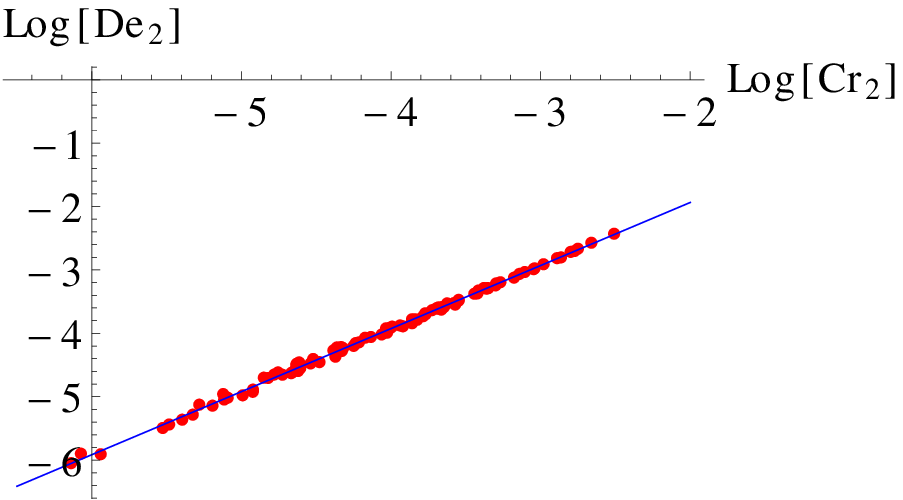}
\includegraphics[width=7.0cm]{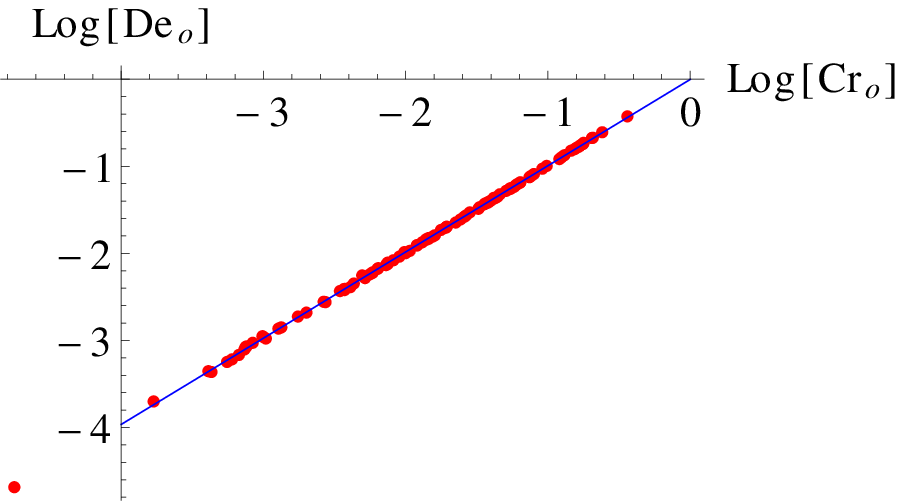}
\end{center}
\caption{
Plots of the $(Cr_1,De_1)$, $(Cr_2,De_2)$ and $(Cr_o,De_o)$
during the fifth computation of the 
correlation coefficients in case of the true function 
$F(g) = (1-\frac{g}{5}+g^2)^{\frac{1}{2}}$.
Here the blue line is the regression line.}
\label{F11-5-CR12}
\end{figure}

%\paragraph{Test of the criteria $(Cr_1, De_1)$}
We check the reliability of the reference quantity $Cr_{1}$ 
by computing the correlation coefficients
between $Cr_1$ and $De_1$.
We compute it five times and the results are
%The following are the correlation coefficients 
%computed five times
%with 
\begin{align}
 \rho^{[1]}_{Cr_1De_1} = 1, \quad
 \rho^{[2]}_{Cr_1De_1} = 0.999998, \quad
 \rho^{[3]}_{Cr_1De_1} = 0.999998, 
\NN \\
 \rho^{[4]}_{Cr_1De_1} = 0.999999, \quad
 \rho^{[5]}_{Cr_1De_1} = 0.999988, 
\label{F11rho1}
\end{align}
where  $\rho^{[i]}_{Cr_1De_1}$ is the 
correlation coefficient computed by the $i$-th calculation.
The plots of $(Cr_1, De_1)$ during the fifth computation 
are shown in Figure~\ref{F11-5-CR12}.
%(The plot for the all examination are 
%in~Fig.\ref{F11CR1DE1}).
Because these are very close to 1, 
these are so strong that we can rely on $Cr_{1}$
as a very good reference quantity for selecting 
a good interpolating function.
Of course the results in \eqref{F11rho1} are 
statistically significant since they are larger than $0.765$.
%these quantities are statistically significant.
%We observe that these are very strong correlation, very close to 
%1, so we can regard $Cr_1$ as a good reference quantity 
%to measure $De_1$
%for the interpolating function for $(1-\frac{g}{5}+g^2)^{\frac{1}{2}}$,
%namely $Cr_1$ is good criteria to investigate 
%which function is better as a interpolating function for
%$(1-\frac{g}{5}+g^2)^{\frac{1}{2}}$.

%\paragraph{Test of the criteria $(Cr_2, De_2)$}
%We will examine also the criteria 
The correlation coefficients between 
$Cr_2$ and $De_2$ are
\begin{align}
 \rho^{[1]}_{Cr_2De_2} = 0.99962083, \quad
 \rho^{[2]}_{Cr_2De_2} = 0.99949903, \quad
 \rho^{[3]}_{Cr_2De_2} = 0.99890653, 
\NN \\
 \rho^{[4]}_{Cr_2De_2} = 0.99916978, \quad
 \rho^{[5]}_{Cr_2De_2} = 0.999553356, 
\label{F11rho2}
% _1
\end{align}
and the ones between $Cr_o$ and $De_o$ are
\begin{align}
 \rho^{[1]}_{Cr_oDe_o} =0.99995961 , \quad
 \rho^{[2]}_{Cr_oDe_o} =0.99996129 , \quad
 \rho^{[3]}_{Cr_oDe_o} =0.99996000 , 
\NN \\
 \rho^{[4]}_{Cr_oDe_o} =0.99996445 , \quad
 \rho^{[5]}_{Cr_oDe_o} =0.99995738 . 
\label{F11rhoo}
%_2
% _1
\end{align}
These quantities are calculated by using the same samples 
%sets of interpolating 
%functions 
as 
\eqref{F11rho1}.
The plots of the $(Cr_2, De_2)$ and $(Cr_o, De_o)$ 
during the fifth computation are listed in 
Figure~\ref{F11-5-CR12}.
%(The plot for the all examination are 
%in~Fig.\ref{F11CR2DE2}).
Since these are also very close to 1, 
$Cr_{2}$ and $Cr_{o}$ are also good reference quantities.
%From the comparison between 
%\eqref{F11rho1} and \eqref{F11rho2},
%$Cr_{1}$ may be a better reference than $Cr_2$.

\subsection{
Functions where one of small-$g$ or large-$g$
expansions is convergent while the other is asymptotic}

\subsubsection{$\varphi^4$ theory}
\label{Sec:phi4-our-cri}
Let us consider the partition function
of the zero-dimensional $\varphi^4$ theory,
\begin{equation}
F(g) = \int^{\infty}_{-\infty} d \varphi \, 
e^{- \frac{\varphi^2}{2} - g^2 \varphi^4}.
\end{equation}
It is well known that the small-$g$ expansion  
is asymptotic while the large-$g$ expansion 
is convergent.
Interpolating functions for this example have been 
studied also in section 4.1
of \cite{Honda:2014bza}.

We assume that we do not know the 
true function $F(g)$. Suppose that 
we know only its large-$g$ and small-$g$
expansions up to only 100-th order.
The expansions are given by 
\begin{equation}
\tilde{F}^{(100)}_{s}(g) = \sum_{k=0}^{100}s_{k} g^k, \qquad
\tilde{F}^{(100)}_{l}(g) = \frac{1}{\sqrt{g}}\sum_{k=0}^{100}l_{k} g^{-k}, 
\label{Eq:Ex2-pre}
\end{equation}
where the coefficients 
$s_{k}$ and $l_{k}$ are already known.
%In the situation only these functions are in hand.

We will show $g_{s}^{\ast},g_{l}^{\ast}, N_{s}^{\ast}$ and $N_{l}^{\ast}$
which were already given in \cite{Honda:2014bza}. 
According to \cite{Honda:2014bza},   
the small-$g$ expansion was clarified to be an asymptotic expansion, 
and its related values are
\begin{equation}
g_{s}^{\ast} = 0.0680628, \qquad N_{s}^{\ast} = 28,
\end{equation}
where the error is $\epsilon = 10^{-7}$.
The large-$g$ expansion was estimated to be 
convergent, and the related quantities are
\begin{equation}
g_{l}^{\ast} = 0.1, \qquad N_{l}^{\ast} = 100.
\end{equation}

By taking into account $N_{s}^{\ast} = 28$ in the small-$g$
expansion, 
we need to redefine the expansions by performing the 
optimal truncation as follows
\begin{equation}
F^{(28)}_{s}(g) = \sum_{k=0}^{28}s_{k} g^k, \qquad
F^{(100)}_{l}(g) = \frac{1}{\sqrt{g}}\sum_{k=0}^{100}l_{k} g^{-k}. 
\label{Eq:Ex2}
\end{equation}
We will construct the interpolating functions based on these expansions
\eqref{Eq:Ex2}. Several interpolating functions are given by eq.~(B.1)
of \cite{Honda:2014bza}. 
%and they are listed in \eqref{Eq:phi4-intpol-list} in 
%Appendix~\ref{app:ex4}.
As in \eqref{Random-linear} and \eqref{Eq:Int-func-conv-1},
we consider %sets of 
the interpolating functions, 
which are linear combinations 
of the functions in eq.~(B.1) of \cite{Honda:2014bza}
%\eqref{Eq:phi4-intpol-list}
with the randomly generated coefficients, 
as the samples.

%with randomly chosen coefficient,
%\begin{align}
%\hat{F}^{[r,s]}(g) 
%=& c_{1}^{[r,s]}F^{(1/2)}_{1,1}(g)
%+c_{2}^{[r,s]}F^{(1/6)}_{1,1}(g)
%+c_{3}^{[r,s]}F^{(1/2)}_{2,2}(g)
%+c_{4}^{[r,s]}F^{(1/6)}_{2,2}(g)
%\NN \\
%&+c_{5}^{[r,s]}F^{(1/10)}_{2,2}(g)
%+c_{6}^{[r,s]}F^{(1/2)}_{3,3}(g)
%+c_{7}^{[r,s]}F^{(1/6)}_{3,3}(g)
%+c_{8}^{[r,s]}F^{(1/14)}_{3,3}(g)
%\NN \\
%&
%+c_{9}^{[r,s]}F^{(1/2)}_{4,4}(g)
%+c_{10}^{[r,s]}F^{(1/6)}_{4,4}(g)
%+c_{11}^{[r,s]}F^{(1/10)}_{4,4}(g)
%+c_{12}^{[r,s]}F^{(1/18)}_{4,4}(g)
%\label{Eq:Int-func-phi4},
%\end{align}
%where 
%\begin{equation}
%\sum_{i=1}^{12}c_{i}^{[r,s]} =  1, \qquad c_{i}^{[r,s]}\ge 0.
%\label{constraint-2}
%\end{equation}
%Meaning of the superscript $[r,s]$ is same as in the previous section 
%\ref{Sec:B-conv-1}.
%\begin{figure}[tbp]
%\begin{center}
%\includegraphics[width=7.0cm]{Fphi-Cr1De1-5.eps}
%\includegraphics[width=7.0cm]{Fphi-Cr2De2-5.eps}
%\end{center}
%\caption{
%Plot of the $(Cr_1,De_1)$ and $(Cr_2,De_2)$
%at the fifth examination of the 
%correlation coefficients in the partition
%function for the zero-dimensional $\varphi^4$ theory case.}
%\label{Fphi-5-Cr12}
%\end{figure}
We calculate the correlation coefficients
by using the samples. %between 
%$Cr_{1}$ and $De_{1}$  as
%well as the ones between $Cr_{2}$ and $De_{2}$.
The correlation coefficients between $Cr_1$  and $De_1$ 
are 
\begin{align}
 \rho^{[1]}_{Cr_1De_1} = 1, \quad
 \rho^{[2]}_{Cr_1De_1} = 0.999823, \quad
 \rho^{[3]}_{Cr_1De_1} = 1, 
\NN \\
 \rho^{[4]}_{Cr_1De_1} = 1, \quad
 \rho^{[5]}_{Cr_1De_1} = 1 .
\label{Fphirho1}
\end{align}
The ones between $Cr_2$ and $De_2$ 
are
\begin{align}
 \rho^{[1]}_{Cr_2De_2} = 0.999999, \quad
 \rho^{[2]}_{Cr_2De_2} = 0.999997, \quad
 \rho^{[3]}_{Cr_2De_2} = 0.999997,
\NN \\
 \rho^{[4]}_{Cr_2De_2} = 0.999997, \quad
 \rho^{[5]}_{Cr_2De_2} = 0.999998,
\label{Fphirho2}
\end{align}
and the ones between $Cr_o$ and $De_o$ 
are
\begin{align}
 \rho^{[1]}_{Cr_oDe_o} = 0.999997, \quad
 \rho^{[2]}_{Cr_oDe_o} = 0.999991, \quad
 \rho^{[3]}_{Cr_oDe_o} = 0.999992,
\NN \\
 \rho^{[4]}_{Cr_oDe_o} = 0.999993, \quad
 \rho^{[5]}_{Cr_oDe_o} = 0.999994.
\label{Fphirhoo}
% _2
\end{align}
Here \eqref{Fphirho2} and \eqref{Fphirhoo} are computed by using
the same samples as \eqref{Fphirho1}.
%{\bf From here modify}
Because these correlation coefficients are almost 1, 
all the $Cr_1$, $Cr_{2}$ and $Cr_{o}$ work very well 
as good reference quantities 
for choosing a good interpolating function.

\subsubsection{Average plaquette in the four-dimensional $SU(3)$ 
pure Yang-Mills theory on the lattice}
\label{Sec:Lat-our-Cri}
As a next example, we consider the average plaquette
\begin{equation}
 P (\beta )
= \Bigl\langle 1
 -\frac{1}{3}{\rm Tr} U_{\mathbf{x} ,\mu} U_{\mathbf{x} +\hat{\mu} ,\nu}   U_{\mathbf{x} +\hat{\nu} ,\mu}^\dag U_{\mathbf{x}  ,\nu}^\dag \Bigr\rangle    
\end{equation}
in the four dimensional $SU(3)$ pure Yang-Mills theory on the 
lattice. 
The action of the theory is given by 
\begin{equation}
 S
=\beta \sum_{\mu <\nu} \sum_{\mathbf{x}} \Bigl[ 1
 -\frac{1}{3} {\rm ReTr} U_{\mathbf{x} ,\mu} U_{\mathbf{x} +\hat{\mu} ,\nu}   U_{\mathbf{x} +\hat{\nu} ,\mu}^\dag U_{\mathbf{x}  ,\nu}^\dag \Bigr] ,
\end{equation}
where 
$U_{\mathbf{x} ,\mu}$ is the link variable 
along the $\mu$-direction at the position $\mathbf{x}$. 
Here $\hat{\mu}$ denotes the unit vector along the $\mu$-direction.
Here the true function for the average plaquette 
$P(\beta_i)$ has been obtained by the Monte Carlo simulation
in \cite{Honda:2014bza}.
\footnote{
The calculations have 
been done at the following values of $\beta$:
$(\beta_1 ,\cdots ,\beta_{29})=$ $(0.1,$ $0.2,$ $0.5,$ $1.0,$ $1.5,$ $2.0,$ $2.5, $ $2.75,$ $3.0,$ $3.25,$ $3.5,$ $3.75,$ $4.0,$ $4.25,$ 
$4.5,$ $4.75,$ $5.0,$ $5.25,$ $5.5,$ $5.75,$ $6.0,$ $6.25,$ $6.5,$ $6.75,$ $7.0,$ $7.5,$ $8.0,$ $9.0,$ $10)$.
} 
The interpolating functions for $P(\beta)$ were also studied in 
\cite{Honda:2014bza}. 

Ref.~\cite{Balian:1974xw}
has given the strong coupling expansion around $\beta =0$, 
\begin{\eq}
P_s^{(15)} (\beta ) = \sum_{k=0}^{15} s_k \beta^k ,
\end{\eq}
where the coefficients are explicitly described in (4.30) 
of \cite{Honda:2014bza}.
%On the other hand, 
The weak coupling expansion around $\beta =\infty$ 
is given by \cite{Bali:2014fea} (see also \cite{DiRenzo:1995qc,DiRenzo:2000ua,Horsley:2012ra,Di Renzo:2004ge})
\begin{\eq}
P_l^{(34)} (\beta ) = \frac{1}{\beta} \sum_{k=0}^{34} l_k \beta^{-k} ,
\end{\eq}
where
the coefficients are listed in (4.32) of \cite{Honda:2014bza}.
%\footnote{
%Actually these values have errors and we are using just their center values.
%See \cite{Bali:2014fea} for details.
%}

According to the study in 
\cite{Honda:2014bza}, 
the small-$\beta$ expansion is convergent and 
its related quantities are
given by \footnote{
In \cite{Osterwalder:1977pc},
the authors have proven that 
the strong coupling expansion in the lattice gauge theory
is convergent. 
%However we should not assert that 
%we have determined 
%the convergent radius precisely as $\beta_{s}^{\ast} = 3.9$. 
%In \cite{Honda:2014bza},
%the ``convergent radius'' $\beta_{s}^{\ast}$ 
%the value of $\beta_{s}^{\ast}$ 
%was given just as a reference value to define the reliable domain.
}
\begin{equation}
\beta_{s}^{\ast} = 3.9, \quad N_{s}^{\ast} = 15.
\end{equation}
On the other hand, the large-$\beta$ expansion turned out to be  
asymptotic, and the related values were given by
%The optimal truncation 
%with error 
%$\epsilon = 10^{-4}$
%is implemented at 
%and the corresponding reliable domain is given by 
\begin{equation}
\beta_{l}^{\ast} = 6.13706,
\quad
N_{l}^{\ast} = 34.
\end{equation}

Several interpolating functions were already given by 
eq.~(B.6) of \cite{Honda:2014bza}. 
%and they are listed 
%in \eqref{app:YM-eq} in Appendix~\ref{app:YM}.
%By using the functions \eqref{app:YM-eq},
As in \eqref{Random-linear} and \eqref{Eq:Int-func-conv-1},
we consider interpolating functions $\hat{P}^{[r,s]}$, 
which are %radomly generated by 
the linear combinations 
of the functions in eq.~(B.6) of \cite{Honda:2014bza} %\eqref{app:YM-eq}
with randomly generated coefficients,
as samples of interpolating functions.
%with randomly chosen coefficients
%$c_{i}^{[r,s]}$,
%\begin{align}
%\hat{P}^{[r,s]}(\beta) 
%=& c_{1}^{[r,s]}P^{(-1)}_{1,1}(\beta)
%+c_{2}^{[r,s]}P^{(-1/3)}_{1,1}(\beta)
%+c_{3}^{[r,s]}P^{(-1)}_{2,2}(\beta)
%+c_{4}^{[r,s]}P^{(-1)}_{3,3}(\beta)
%\NN \\
%&+c_{5}^{[r,s]}P^{(-1)}_{4,4}(\beta)
%+c_{6}^{[r,s]}P^{(-1)}_{5,5}(\beta)
%+c_{7}^{[r,s]}P^{(-1)}_{6,6}(\beta)
%+c_{8}^{[r,s]}P^{(-1)}_{7,7}(\beta)
%\NN \\
%&+c_{9}^{[r,s]}P^{(-1)}_{8,8}(\beta)
%+c_{10}^{[r,s]}P^{(-1)}_{9,9}(\beta)
%+c_{11}^{[r,s]}P^{(-1)}_{10,10}(\beta)
%+c_{12}^{[r,s]}P^{(-1)}_{11,11}(\beta)
%\NN \\
%&+c_{13}^{[r,s]}P^{(-1)}_{12,12}(\beta)
%+c_{14}^{[r,s]}P^{(-1)}_{13,13}(\beta)
%+c_{15}^{[r,s]}P^{(-1)}_{14,14}(\beta)
%+c_{16}^{[r,s]}P^{(-1)}_{15,15}(\beta),
%\label{Eq:Int-func-lattice}
%\end{align}
%where 
%\begin{equation}
%\sum_{i=1}^{16}c_{i}^{[r,s]} =  1, \qquad c_{i}^{[r,s]}\ge 0.
%\label{constraint-lattice}
%\end{equation}

In terms of the true function $P(\beta_i)$, 
the sets $(Cr_{1}, De_{1})$  $(Cr_{2}, De_{2})$ and $(Cr_{o},De_{o})$
are given by
\begin{align}
 De_{1} = {\rm Max}\left\{\left|P(\beta_{i})-\hat{P}^{[r,s]}(\beta_i)\right|;
i = 1\sim 29
\right\},
\label{DE1lat}
\\
 De_{2} = \frac{1}{29} \sum_{i=1}^{29} \left|
\frac{P(\beta_i)-\hat{P}^{[r,s]}(\beta_i)}{P(\beta_i)}
\right|,
\label{DE2lat}
\\
 De_{o} = \frac{1}{29} \sum_{i=1}^{29} \left|
P(\beta_i)-\hat{P}^{[r,s]}(\beta_i)
\right|,
\label{DEolat}
\end{align}
and 
\begin{equation}
Cr_1 = {\rm Max}\left[
\left\{
\left|\hat{P}^{[r,s]}(\beta) - P_{s}^{(15)}(\beta)
\right|
; 0 \le \beta \le \beta_{s}^{\ast}
\right\}\cup 
\left\{\left|\hat{P}^{[r,s]}(\beta) - P_{l}^{(34)}(\beta)
\right|
;  \beta_{l}^{\ast} \le \beta 
\right\}
\right],
\label{CR12lat}
\end{equation}
\begin{equation}
Cr_2 = \int^{\beta_{s}^{\ast}}_{0} d\beta
\left|\frac{\hat{P}^{[r,s]}(\beta) - P_{s}^{(15)}(\beta)}{P_{s}^{(15)}(\beta)}
\right|
+
\int_{\beta_{l}^{\ast}}^{\infty} d\beta
\left|
\frac{\hat{P}^{[r,s]}(\beta) - P_{l}^{(34)}(\beta)}{P_{l}^{(34)}(\beta)}
\right|,
\label{CR22lat}
\end{equation}
\begin{equation}
Cr_o = \int^{\beta_{s}^{\ast}}_{0} d\beta
\left|\hat{P}^{[r,s]}(\beta) - P_{s}^{(15)}(\beta)
\right|
+
\int_{\beta_{l}^{\ast}}^{\infty} d\beta
\left|
\hat{P}^{[r,s]}(\beta) - P_{l}^{(34)}(\beta)
\right|.
\label{CRo2lat}
\end{equation}
\begin{figure}[tbp]
\begin{center}
\includegraphics[width=7.0cm]{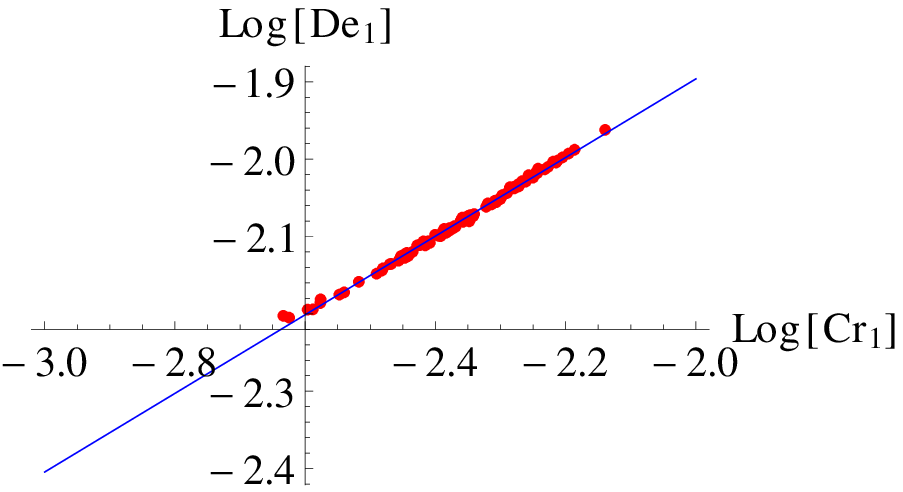}
\includegraphics[width=7.0cm]{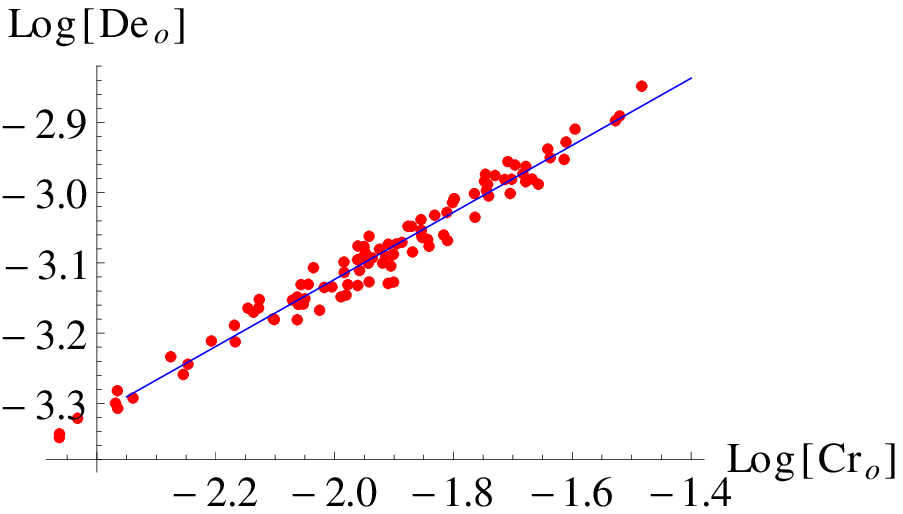}
\includegraphics[width=7.0cm]{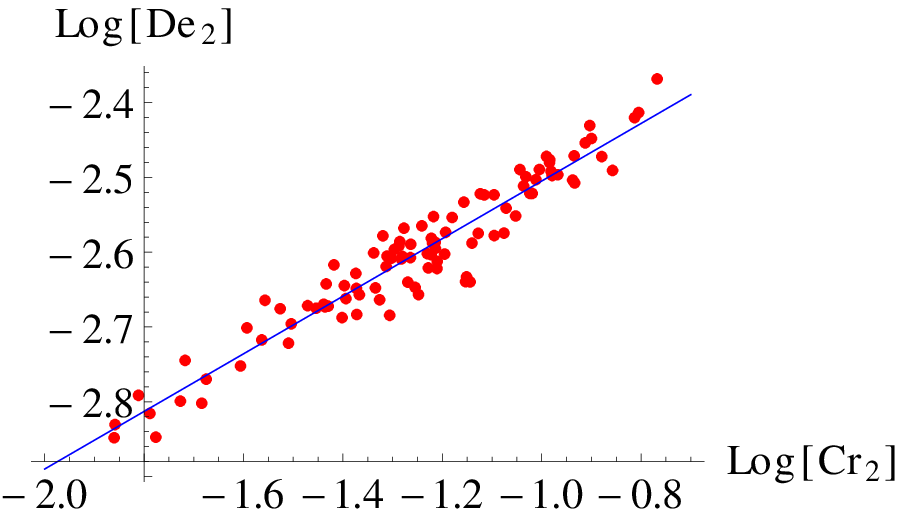}
\end{center}
\caption{
Plots of $(Cr_1,De_1)$, $(Cr_o,De_o)$ and $(Cr_2,De_2)$
during the 5-th calculation of the 
correlation coefficients in case of the 
average plaquette in 
the 4-dimensional 
$SU(3)$ pure Yang-Mills theory on the lattice.
We can see that the correlation between $Cr_1$ and $De_1$ 
is the strongest.
}
\label{Flat-5-Cr12}
\end{figure}

Let us check whether $Cr_{1,2,o}$ work well as  
reference quantities or not by 
computing the correlation coefficients between $Cr_{j}$ and $De_{j}$
for $j \in \{1,2,o\}$.
The correlation coefficients between 
$Cr_{1}$ and $De_{1}$ are
\begin{align}
 \rho^{[1]}_{Cr_1De_1} = 0.998887, \quad
 \rho^{[2]}_{Cr_1De_1} = 0.998455, \quad
 \rho^{[3]}_{Cr_1De_1} = 0.998134, 
\NN \\
 \rho^{[4]}_{Cr_1De_1} = 0.999318, \quad
 \rho^{[5]}_{Cr_1De_1} = 0.998927.
\label{Cr1De1-lattice}
\end{align}
The ones between $Cr_o$ and $De_o$  are
\begin{align}
 \rho^{[1]}_{Cr_oDe_o} = 0.990178, \quad
 \rho^{[2]}_{Cr_oDe_o} = 0.987834, \quad
 \rho^{[3]}_{Cr_oDe_o} = 0.975302, 
\NN \\
 \rho^{[4]}_{Cr_oDe_o} = 0.982322, \quad
 \rho^{[5]}_{Cr_oDe_o} = 0.982845,
\label{CroDeo-lattice}
% _2
\end{align}
%While 
and the correlation coefficients between $Cr_2$ and $De_2$  are
\begin{align}
 \rho^{[1]}_{Cr_2De_2} = 0.98446, \quad
 \rho^{[2]}_{Cr_2De_2} = 0.968163, \quad
 \rho^{[3]}_{Cr_2De_2} = 0.938417, 
\NN \\
 \rho^{[4]}_{Cr_2De_2} = 0.959227, \quad
 \rho^{[5]}_{Cr_2De_2} = 0.956638,
\label{Cr2De2-lattice}
% _1
\end{align}
where \eqref{CroDeo-lattice} and 
\eqref{Cr2De2-lattice} are computed by using 
the same samples as
\eqref{Cr1De1-lattice}.
These are of course statistically significant, 
and these are very close to 1.
So $Cr_{1,2,o}$ are sufficiently good reference quantities for
selecting a good interpolating function. 
By comparing \eqref{Cr1De1-lattice} with \eqref{CroDeo-lattice} and 
\eqref{Cr2De2-lattice},
$Cr_{1}$ turns out to be the best reference quantity because of the strongest
correlation to the actual degree of deviation.
(We can also see it by comparing the plots %of 
%$(Cr_1,De_1)$ 
%with the ones of $(Cr_2,De_2)$ and $(Cr_o,De_o)$ 
in Figs.~\ref{Flat-5-Cr12})

\subsection{Functions where the both expansions are asymptotic}
\subsubsection{$F(g)= e^{\frac{1}{g^4}}e^{ g^4}
K_{\frac{1}{4}}\left(\frac{1}{g^4}\right)K_{\frac{1}{4}}(g^4)$}
\label{Sec:Basym-our-cri}
\begin{figure}[t]
\begin{center}
\includegraphics[width=6.5cm]{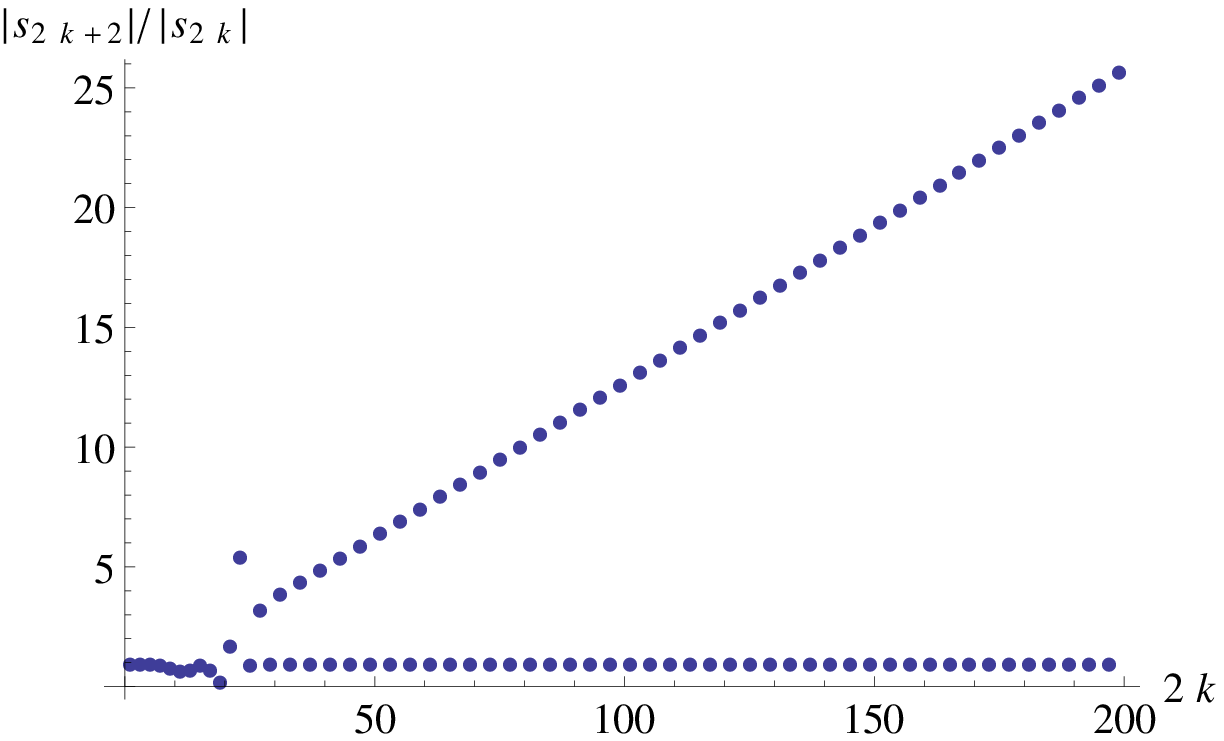}
\includegraphics[width=6.5cm]{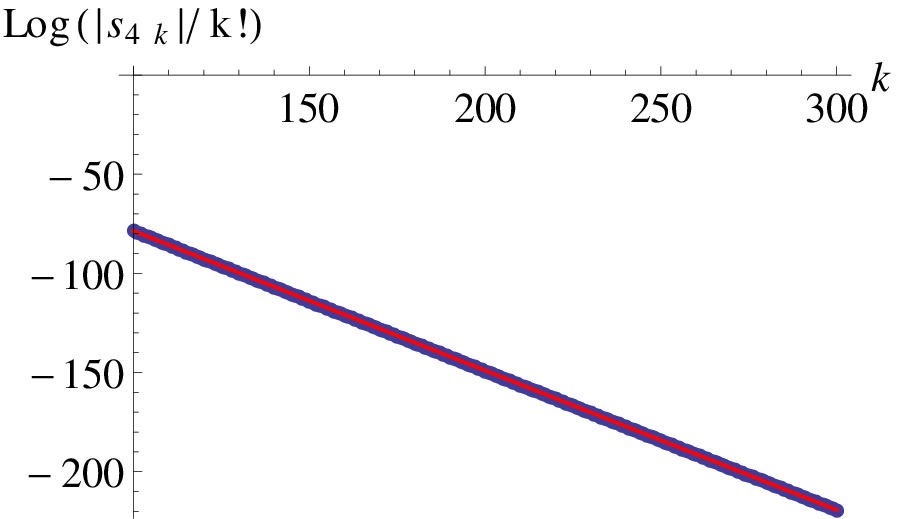}
\includegraphics[width=6.5cm]{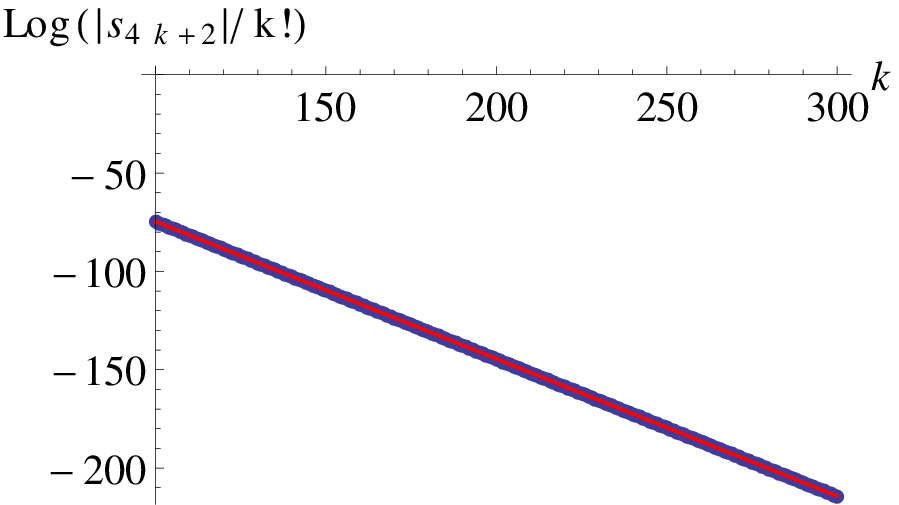}
\end{center}
\caption{
[Top Left] $|s_{2k+2}|/|s_{2k}|$ v.s $k$
[Top Right] $\log (|s_{4k}/k!|)$ v.s $k$
[Bottom] $\log (|s_{4k+2}/k!|)$ v.s $k$.}
\label{Fig:plot-both-1}
\end{figure}

We will consider the example with following true function
\begin{equation}
F(g)= e^{\frac{1}{g^4}}e^{ g^4}
K_{\frac{1}{4}}\left(\frac{1}{g^4}\right)
K_{\frac{1}{4}}(g^4), 
\end{equation}
where $K_{\frac{1}{4}}(x)$ is the modified Bessel function of the second kind
with order $\frac{1}{4}$.
The small-$g$ and large-$g$ expansions of the function take
following forms
\begin{align}
F^{(N_s)}_{s}(g) =& g \sum_{k = 0}^{{\rm floor}(\frac{N_s}{2})} s_{2k} g^{2k}
= g \sum_{k = 0}^{{\rm floor}(\frac{N_s}{4})} s_{4k} g^{4k}
+g \sum_{k = 0}^{{\rm floor}(\frac{N_s}{4})} s_{4k+2} g^{4k+2},
\\
F^{(N_l)}_{l}(g) =&  \frac{1}{g} \sum_{k = 0}^{{\rm floor}(\frac{N_l}{2})} 
l_{2k} g^{-2k}
= \frac{1}{g} \sum_{k = 0}^{{\rm floor}(\frac{N_l}{4})} l_{4k} g^{-4k}
+ \frac{1}{g} \sum_{k = 0}^{{\rm floor}(\frac{N_l}{4})} l_{4k+2} g^{-4k-2}.
\end{align}
Here ${\rm floor}(x) = \max \{n \in \Z ; n \le x\}$
which is called as the floor function. 
The $F(g)$ is a symmetric function under the exchange of $g$ and $1/g$.
By extrapolating the data of $\left|\frac{s_{2k+2}}{s_{2k}}\right|$ 
in [Top Left] of Figs.~\ref{Fig:plot-both-1}, 
the small-$g$ expansion turns out to be asymptotic 
since the ratio diverges in large $k$.
Analogously we can see that the large-$g$ expansion is also
asymptotic.
We observe that 
$\left|\frac{s_{4k+4}}{s_{4k}}\right|$,
$\left|\frac{s_{4k+6}}{s_{4k+2}}\right|$,
$\left|\frac{l_{4k+4}}{l_{4k}}\right|$ and  
$\left|\frac{l_{4k+6}}{l_{4k+2}}\right|$ behave as
linear in $k$, so we put ansatz 
\begin{equation}
s_{4k} \sim c_1 A_{1}^{k} k!, \qquad s_{4k+2} \sim c_2 A_{2}^{k} k!,
\qquad l_{4k} \sim d_1 B_{1}^{k} k!, \qquad l_{4k+2} \sim d_2 B_{2}^{k} k!,
\end{equation}
in large $k$.
By fitting the log-plots as in [Top Right] and [Bottom] of 
Figs.~\ref{Fig:plot-both-1}, we obtain
\begin{align}
&c_{1} = d_{1} = 0.00027207, \quad A_{1} = B_{1} = 
0.494761,
\qquad 
\nonumber \\
&c_{2} = d_{2} = 0.00853819, \quad A_{2} = B_{2} =0.497397.
\end{align}
Because of $A_{1}\sim B_{1} \sim A_{2} \sim B_{2} \sim \frac{1}{2}$,
large order terms in the small-$g$ 
expansion should be 
\begin{equation}
\pm g (c_1 - c_2 g^2)\left(\frac{1}{2}\right)^{k}
k! g^{4k} \label{eq:k-term}
\end{equation}
where $k \gg 1$.
The relative minus sign inside $(c_1 - c_2 g^2)$ 
is required 
because of our observation that 
$s_{4k}\times s_{4k+2} < 0$
upto $k = 100$.
We expect that the optimal truncation will be implemented 
at order less than $100$.
Based on \eqref{eq:k-term},
by following the procedure in Appendix \ref{Sec:Optimize},
we can obtain
$N_s^{\ast}$ as well as $g_{s}^{\ast}$.
According to the procedure, 
\begin{equation}
N_{s}^{\ast} = 4\left|\frac{1}{A{g_{s}^{\ast}}^4}\right|.
\end{equation}
$g_{s}^{\ast}$ is given by the solution of the 
following equation
\begin{equation}
\log \epsilon =  \log g_{s}^{\ast} + \log |c_{1}-c_{2}{g_{s}^{\ast}}^2| 
- \frac{2}{{g_{s}^{\ast}}^{4}}.
\end{equation}
By solving the above
(here we set $\epsilon = 10^{-7}$),
we obtain
\begin{equation}
g_{s}^{\ast}  = 0.6676, 
\quad N_{s}^{\ast} = 40.
\end{equation}
%We perform the analogous discussion
Also in the large-$g$ expansion,
by following the analogous procedure, 
we obtain 
\begin{equation}
g_{l}^{\ast} = 1.4979 , \quad N_{l}^{\ast} = 40.
\end{equation}
Based on the expansions
$F^{(N_s^{\ast})}_{s}(g), F^{(N_l^{\ast})}_{l}(g)$
with $N^{\ast}_{s} = N_{l}^{\ast} = 40$, 
we construct the interpolating functions $F_{m,n}^{(\alpha)}$ as
\begin{equation}
F_{m,n}^{(\alpha)}(g)
= \frac{\sqrt{\pi}\Gamma(\frac{1}{4})}{2^{\frac{5}{4}}}g
\left(\frac{1+ \sum_{k = 1}^{p} c_{k}g^{2k}}{1+ \sum_{k = 1}^{q} d_{k}g^{2k}}
\right)^{\alpha},
\end{equation}
where 
\begin{equation}
p = \frac{1}{2}\left(
\frac{1}{2}m + \frac{1}{2}n+1 - \frac{1}{\alpha}
\right), \quad
q = \frac{1}{2}\left(
\frac{1}{2}m + \frac{1}{2}n+1 + \frac{1}{\alpha}
\right),
\end{equation}
\begin{equation}
\left|\frac{2}{m+n+2} \right| \le \left| \alpha\right|,
\qquad
\alpha = 
\left\{ \begin{matrix}
\frac{1}{2 \ell +1} & {\rm for} \quad & \frac{1}{2}(m+n): {\rm even}
\\
\frac{1}{2 \ell } & {\rm for} \quad & \frac{1}{2}(m+n): {\rm odd}
\end{matrix} \right. , \qquad\ell \in \mathbb{Z}.
\end{equation}
Explicit forms of these are described 
in \eqref{List-Int-Asym}
in the Appendix.~\ref{Eq:Ex-Both}.
%\begin{figure}[t]
%\begin{center}
%\includegraphics[width=7.4cm]{3-func-plot-both-asym.eps}
%\includegraphics[width=7.4cm]{diff-both-asym2.eps}
%\includegraphics[width=3.4cm]{plot2_newlattice.eps}
%\includegraphics[width=3.4cm]{plot3_newlattice.eps}
%\includegraphics[width=6.4cm]{plot4_newlattice.eps}
%\includegraphics[width=6.4cm]{plot5_newlattice.eps}
%\end{center}
%\caption{
%[Left] 
%Plot of the function
%$F(g)= e^{\frac{1}{g^4}}e^{ g^4}
%K_{\frac{1}{4}}(\frac{1}{g^4})
%K_{\frac{1}{4}}(g^4)$ 
%(Blue solid line)
%and the  interpolating functions $F^{(1)}_{2,2}$
%(Red dotted line) and $F^{(1/3)}_{2,2}$
%(Black dashed line). 
%These values are so close that they overlap each others
%in this graph.
%[Right] Differences between the interpolating functions and $F(g)$ normalized
%by $F(g)$.
%We can see that the interpolating function is so close to 
%$F(g)$ that deviations are less than $0.4 \%$ of the $F(g)$.}
%\label{Fig:plot-both-3}
%$\end{figure}

As in \eqref{Random-linear} and \eqref{Eq:Int-func-conv-1},
we consider randomly generated 
linear combinations 
of the functions \eqref{List-Int-Asym}
as the samples of the interpolating functions.
%chosen by the random sampling.
%\begin{align}
%\hat{F}^{[r,s]}(g) 
%=& c_{1}^{[r,s]}F^{(1)}_{2,2}(g)
%+c_{2}^{[r,s]}F^{(1/3)}_{2,2}(g)
%+c_{3}^{[r,s]}F^{(1/4)}_{2,4}(g)
%+c_{4}^{[r,s]}F^{(1/5)}_{2,6}(g)
%\NN \\
%&+c_{5}^{[r,s]}F^{(1)}_{4,4}(g)
%+c_{6}^{[r,s]}F^{(1/3)}_{4,4}(g)
%+c_{7}^{[r,s]}F^{(1/5)}_{4,4}(g)
%+c_{8}^{[r,s]}F^{(1/2)}_{4,6}(g)
%\NN \\
%&
%+c_{9}^{[r,s]}F^{(1/6)}_{4,6}(g)
%+c_{10}^{[r,s]}F^{(1/7)}_{4,8}(g)
%+c_{11}^{[r,s]}F^{(1/5)}_{6,6}(g)
%+c_{12}^{[r,s]}F^{(1/7)}_{6,6}(g)
%\NN \\
%&+c_{13}^{[r,s]}F^{(1/9)}_{8,8}(g)
%\label{Eq:Int-func-Basym}
%\end{align}
%where 
%\begin{equation}
%\sum_{i=1}^{13}c_{i}^{[r,s]} =  1, \qquad c_{i}^{[r,s]}\ge 0.
%\label{constraint-4}
%\end{equation}
\begin{figure}[tbp]
\begin{center}
\includegraphics[width=7.0cm]{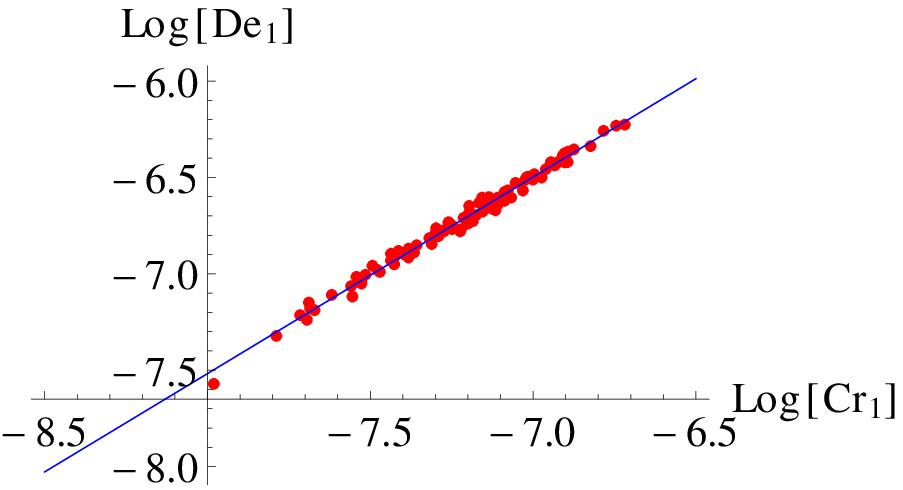}
\includegraphics[width=7.0cm]{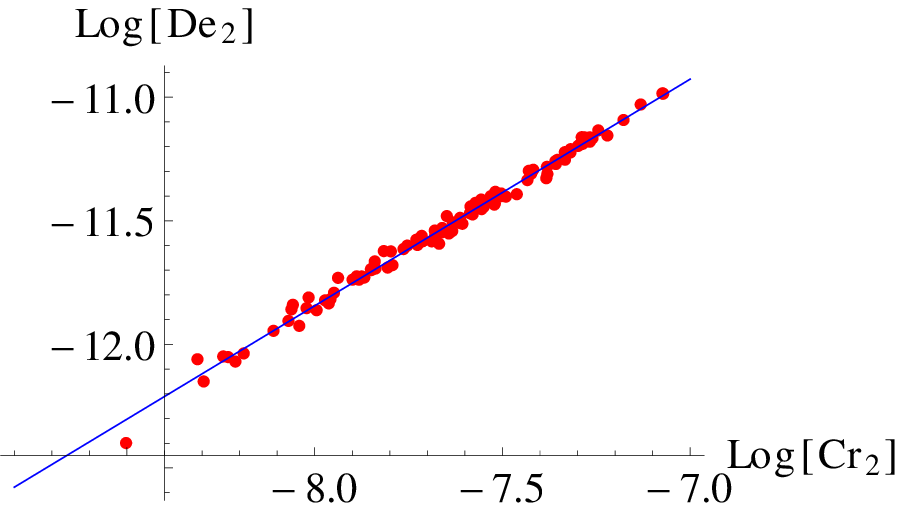}
\includegraphics[width=7.0cm]{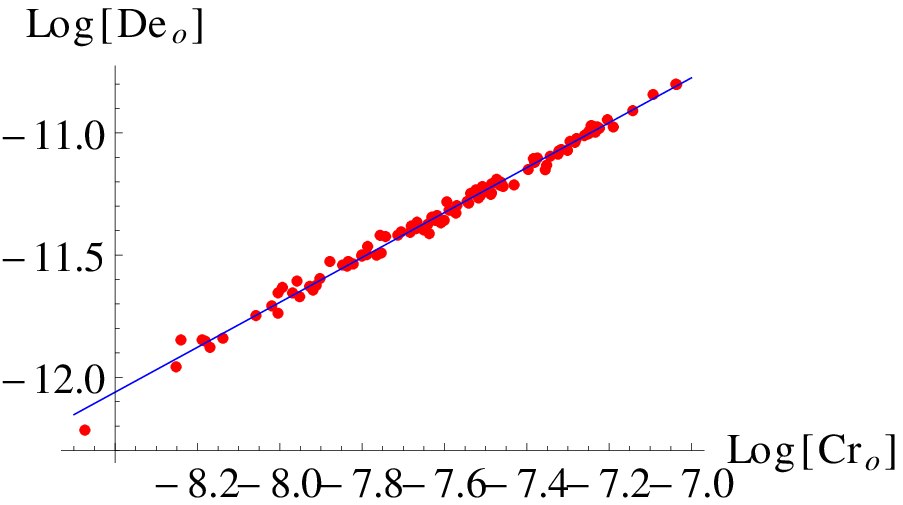}
\end{center}
\caption{
Plots of the $(Cr_1,De_1)$, $(Cr_2,De_2)$
and $(Cr_o,De_o)$
during the 5-th calculation of the 
correlation coefficients in case of the 
$F(g)= e^{\frac{1}{g^4}}e^{ g^4}
K_{\frac{1}{4}}(\frac{1}{g^4})
K_{\frac{1}{4}}(g^4)$
}
\label{Fasym-5-Cr12}
\end{figure}
We check the reference quantities
by computing the correlation coefficients. %between 
%$Cr_{1}$ and $De_{1}$ and the one between $Cr_{2}$ and $De_{2}$.
The correlation coefficients between $Cr_{1}$ and $De_{1}$ are
\begin{align}
 \rho^{[1]}_{Cr_1De_1} = 0.98965, \quad
 \rho^{[2]}_{Cr_1De_1} = 0.992231, \quad
 \rho^{[3]}_{Cr_1De_1} = 0.994823, 
\NN \\
 \rho^{[4]}_{Cr_1De_1} = 0.991528, \quad
 \rho^{[5]}_{Cr_1De_1} = 0.994911.
\label{FBasymrho1}
\end{align}
The ones between $Cr_{2}$ and $De_{2}$ are
\begin{align}
 \rho^{[1]}_{Cr_2De_2} = 0.996086, \quad
 \rho^{[2]}_{Cr_2De_2} = 0.996022, \quad
 \rho^{[3]}_{Cr_2De_2} = 0.996846, 
\NN \\
 \rho^{[4]}_{Cr_2De_2} = 0.995829, \quad
 \rho^{[5]}_{Cr_2De_2} = 0.996668,
\label{FBasymrho2}
% _1
\end{align}
and the ones between $Cr_{o}$ and $De_{o}$ are
\begin{align}
 \rho^{[1]}_{Cr_oDe_o} = 0.996176, \quad
 \rho^{[2]}_{Cr_oDe_o} = 0.99628, \quad
 \rho^{[3]}_{Cr_oDe_o} = 0.996872, 
\NN \\
 \rho^{[4]}_{Cr_oDe_o} = 0.995994, \quad
 \rho^{[5]}_{Cr_oDe_o} = 0.996741.
\label{FBasymrhoo}
% _2
\end{align}
In the computation \eqref{FBasymrho2} and 
\eqref{FBasymrhoo}, 
we use the same samples as \eqref{FBasymrho1}. 
Because these are exceeding 0.99 
(of course statistically significant),
we can take all $Cr_{1}, Cr_{2}$ and $Cr_{o}$ as good reference quantities
for selecting a good interpolating function.
In this case, $Cr_{2,o}$ are slightly 
better than $Cr_{1}$ but the differences are not so significant
according to the comparison between their plots
in Figure~\ref{Fasym-5-Cr12}.

\subsection{Functions with sharp peak outside the reliable domain }
\begin{figure}[tbp]
  \begin{center}
  \subfigure[]{\includegraphics[width=5.0cm]{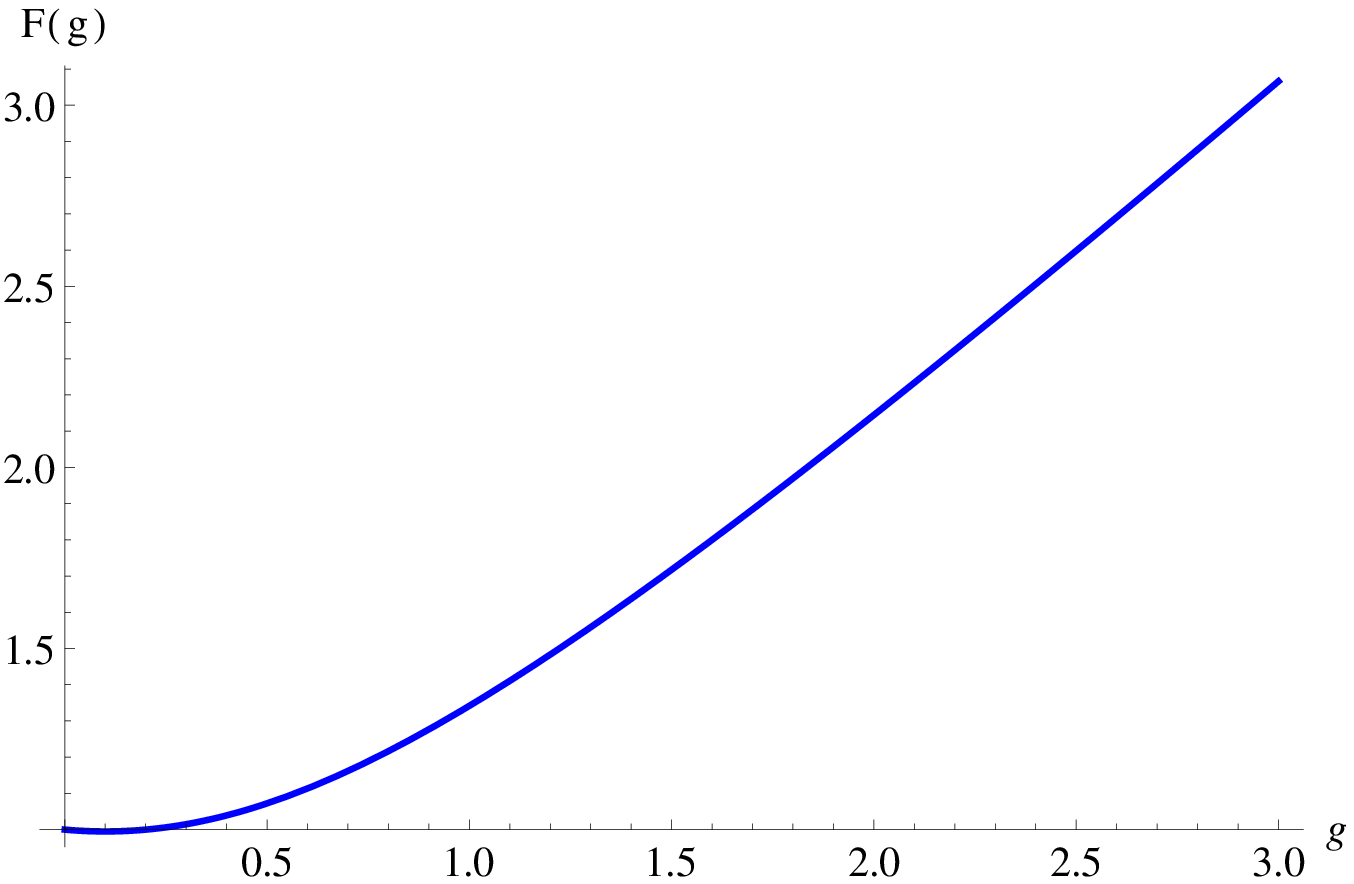}
  }
  \subfigure[]{\includegraphics[width=5.0cm]{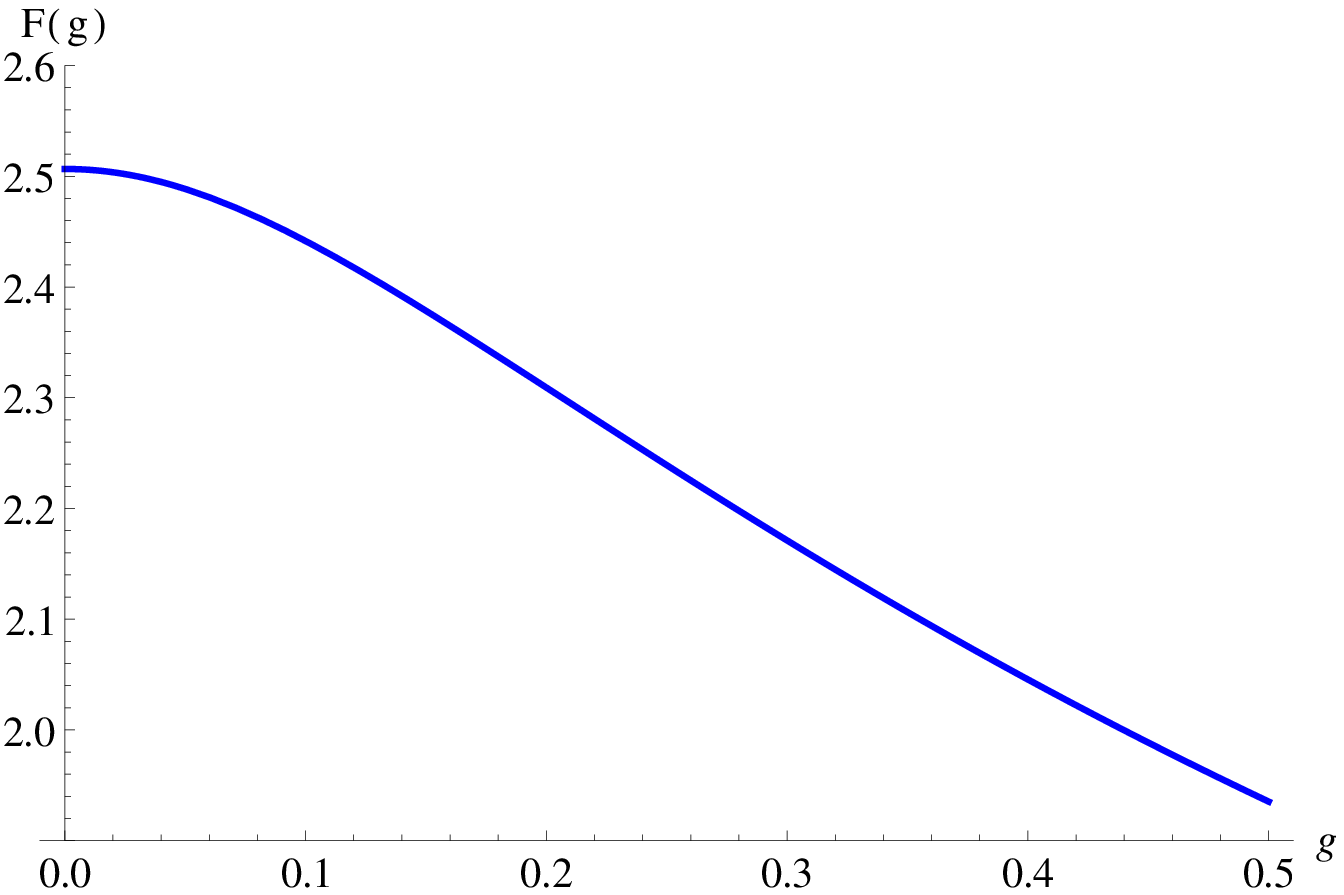}
  }
\\
  \subfigure[]{\includegraphics[width=5.0cm]{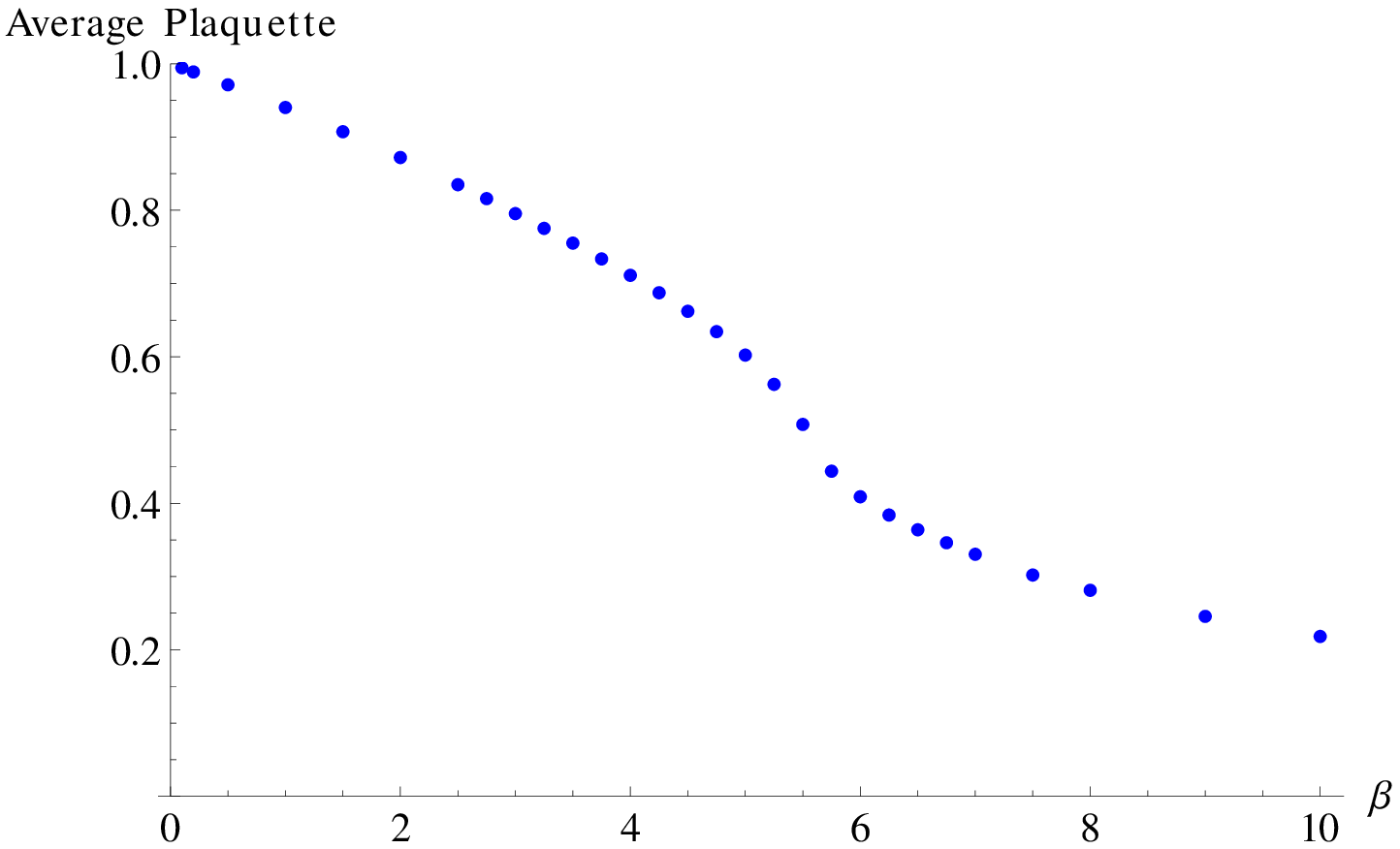}
}
  \subfigure[]{\includegraphics[width=5.0cm]{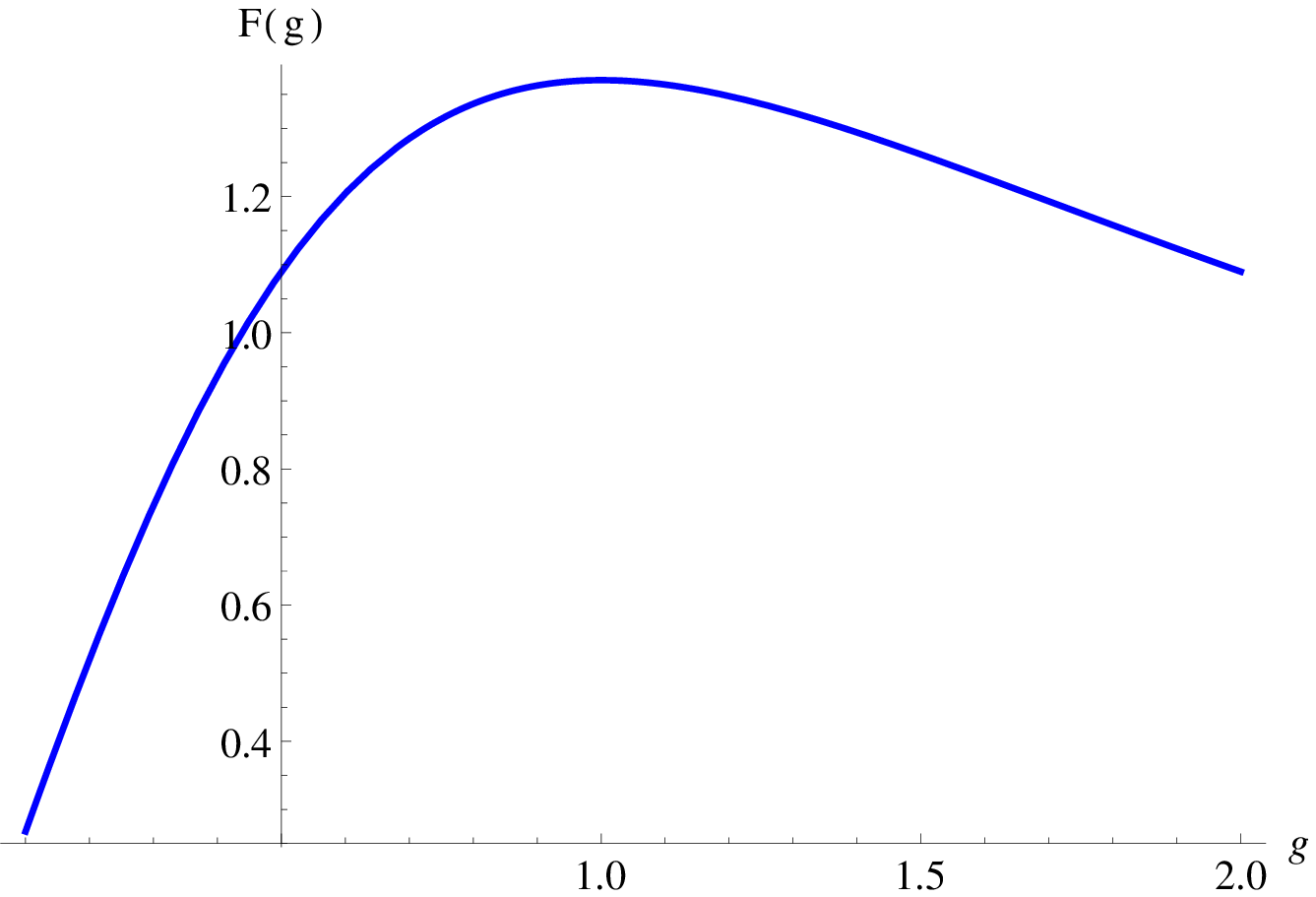}
}
  \end{center}
  \vspace{-0.5cm}
  \caption{(a) Plot of the function $F(g) = (1-\frac{g}{5}+g^2)^\frac{1}{2}$
to $g$,
(b) Plot of $F(g) = \int^{\infty}_{-\infty} d \varphi
e^{- \frac{\varphi^2}{2} - g^2 \varphi^4}$ to $g$,
(c) Plots of Average plaquette in the $SU(3)$ pure Yang-Mills theory on the lattice to $\beta$,
(d) Plot of the function $F(g)= e^{\frac{1}{g^4}}e^{ g^4}
K_{\frac{1}{4}}(\frac{1}{g^4})K_{\frac{1}{4}}(g^4)$ to $g$.}
\label{soft-graph}
 \end{figure}

So far we have checked
$Cr_{1}$, $Cr_{2}$ and $Cr_{o}$ for the functions 
which do not have the sharp peak
outside the reliable domain
(see Figs.~\ref{soft-graph}).
For such functions, all $Cr_{1,2,o}$ 
seem to be good reference quantities,
and $Cr_{1}$ seems to be the best.

\begin{figure}[t]
\begin{center}
\includegraphics[width=9.0cm]{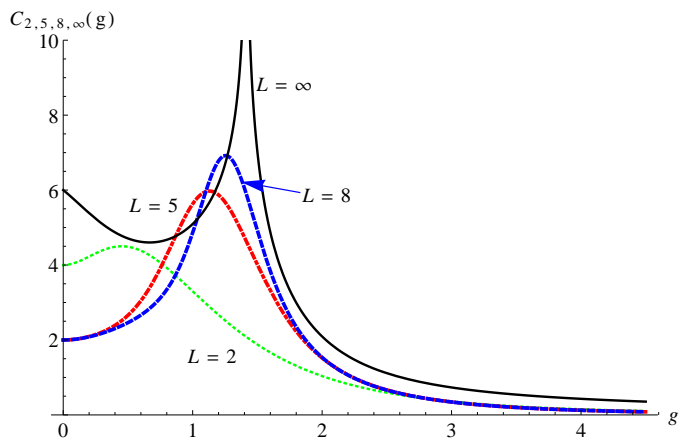}
\end{center}
\caption{
We plot the specific heat 
in the 2 dimensional Ising model against $g$ for various lattice sizes
$L = 2,5,8,\infty$. 
In cases of $L=5,8$, the peaks locate outside the 
reliable domain, where $g_{s}^{\ast} = 0.4$, $g_{l}^{\ast} = 3.8$
in $L = 5$, and $g_{s}^{\ast} = 0.4$, $g_{l}^{\ast} = 3.7$
in $L = 8$ case.}
\label{fig:specific}
\end{figure}

In this subsection, we will investigate the 
%$Cr_{1},Cr_{2}$ and $Cr_{o}$ 
reference quantities
for the functions 
with sharp peak outside the reliable domain.
%We expect that the $Cr_{1}$ will not work well
%while the others $Cr_{2,o}$ keep working as a reference quantity.
We will check 
whether they work well or not even in the presence of the sharp peak, 
by calculating the correlation coefficients.
In this subsection,
as the functions with sharp peak,
we use the specific heat functions 
in the two dimensional 
Ising model with lattice size $L = 5,8, \infty$.
By Figure~\ref{fig:specific}, we can see that the functions 
with $L = 5,8$ have sharp peak moreover the 
$L = \infty$ function has the singular point which is regarded as the 
phase transition point.
\footnote{We have also studied the $L=2$ case.
In this case, there are not sharp peaks as we can see in 
Figure~\ref{fig:specific}.
We have checked that all the $Cr_{1,2,o}$ work well as reference 
quantities, and $Cr_{1}$ is the best. 
%where 
%$Cr_{o}$ is the reference quantity in \cite{Honda:2014bza}.
The discussion on the $L=2$ case belongs to the same class as 
subsubsection~\ref{Sec:B-conv-1}, 
because both the small-$g$ and large-$g$ expansions are convergent.}

\paragraph{Concise review of the specific heat in the two-dimensional Ising 
model}
We will consider the two-dimensional Ising model 
on the $L\times L$ square lattice with periodic boundary condition.
The detailed explanation is in section 4.2 of \cite{Honda:2014bza}.
The Hamiltonian of the Ising model is described as
\begin{\eq}
H = -J \sum_{\langle \mathbf{x},\mathbf{y} \rangle} 
\sigma_{\mathbf{x}} \sigma_{\mathbf{y}}.
\label{eq:Ising-Hamiltonian}
\end{\eq}
Here
$\mathbf{x}$ and $\mathbf{y}$ 
denote the locations of the lattice sites taking integer value.
The notation 
$\langle \mathbf{x},\mathbf{y} \rangle$
indicates the sum over pairs of adjacent spins.
$\sigma_{\mathbf{x}}$ describes
the spin variable at $\mathbf{x}$ taking 
$\sigma_{\mathbf{x}} = \pm 1$.
$J$ denotes the exchange energy (coupling constant)
between the nearest neighbor spins.
The partition function of the model 
with respect to the temperature $T$
is 
\begin{\eq}
Z_L (K) = \sum_{\{ \rm state \}} e^{-\frac{1}{T} H} 
= \sum_{\{ \rm state \}} e^{K \sum_{(\mathbf{x},\mathbf{y})} \sigma_{\mathbf{x}} \sigma_{\mathbf{y}}}  ,\quad
{\rm with}\ K=\frac{J}{T},
\end{\eq}
%\cite{Kastening} gave the exact 
%solution of the partition function,
and it has been calculated exactly in \cite{Kastening},
\begin{\eqa}
&&Z_L (K) = \frac{1}{2} \left( S_{11} (K)  +2S_{10} (K) - S_{00} (K)  \right) ,\NN\\
&& S_{\sigma_1 \sigma_2} (K) = 2^{L^2} \prod_{p,q=0}^{L-1} \Bigl[ 
\cosh^2{(2K)} -\sinh{(2K)} \Bigl( \cos{\frac{(2p+\sigma_1 )\pi}{L}} +\cos{\frac{(2q+\sigma_2 )\pi}{L}} \Bigr)
\Bigr]^{\frac{1}{2}}.
\NN \\
\end{\eqa}
Based on the partition function, 
we introduce the following quantity 
\begin{\eq}
C_L (K) =  \frac{1}{L^2} \frac{\partial^2}{\partial K^2} \log{Z_L (K)}.
\label{eq:CL}
\end{\eq}
In terms of the $C_L(K)$, 
the specific heat can be given by $K^2 C_L (K)$.
For considering the power series 
form \eqref{eq:asymptotics} of 
both the high and low temperature expansion, 
%In order to transform the both expansions 
%to the power series form \eqref{eq:asymptotics},
we introduce the parameter $g$ by
\begin{\eq}
e^{2K} = 1+g.
\end{\eq}
Then the power series expansion of $g$
around $g = 0$ corresponds to the high temperature expansion
while the $1/g$ expansion around the 
$g = \infty$ corresponds to the low temperature expansion.

\subsubsection{$5 \times 5$ lattice}
\label{Sec:L5-1}
%Let us study $Cr_1$ and $Cr_2$ for the interpolating functions
%of $C_{5}(g)$ which is the function for the specific heat in the 
%$L=5$, which is obtained by substituting $L=5$ into \eqref{eq:CL}. 
For $L=5$, the specific heat $C_5(g)$ is obtained by substituting 
$L=5$ into \eqref{eq:CL}.
The true function $C_{5}(g)$ has 
a sharp peak as shown in Figure~\ref{fig:specific}.
We will pay attention to how the peak affects the correlation
coefficients $\rho_{Cr_1De_1}$,
$\rho_{Cr_2De_2}$ and $\rho_{Cr_oDe_o}$.

\begin{figure}[t]
\begin{center}
\includegraphics[width=7.0cm]{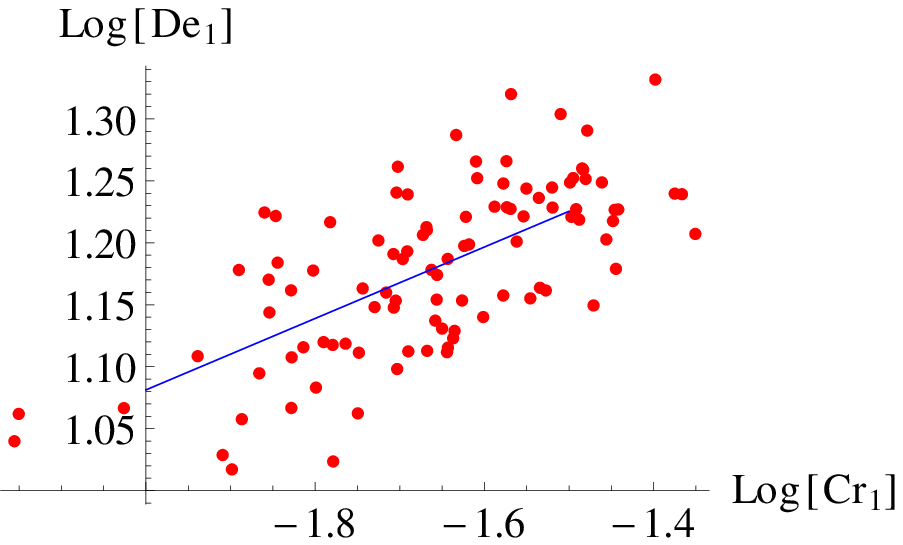}
\includegraphics[width=7.0cm]{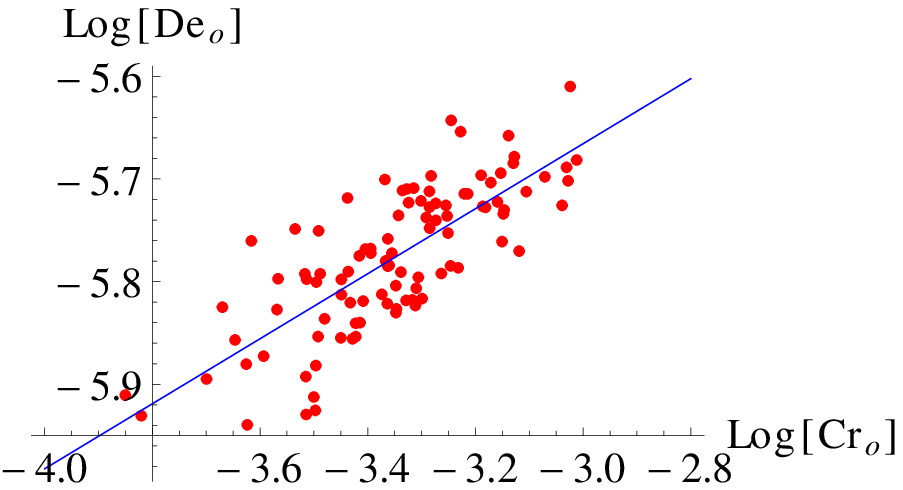}
\includegraphics[width=7.0cm]{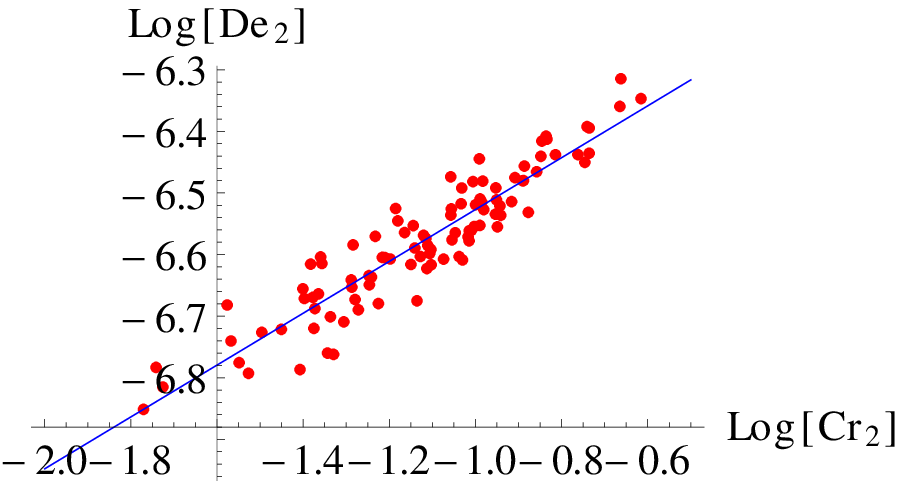}
\end{center}
\caption{Plots of 
$(Cr_1,De_1)$,
$(Cr_o,De_o)$ 
and $(Cr_2,De_2)$ 
during the 5-th computation of the correlation coefficients
in case of the $L=5$ Ising model.
We can see that the set $(Cr_{2}, De_{2})$ have much stronger correlation 
than the others, where the plots of $(Cr_{o}, De_{o})$ and $(Cr_{1}, De_{1})$ 
are more scattered.}
\label{fig:L5Cr1}
\end{figure}

We assume that 
we know only the small-$g$ and large-$g$ expansions 
up to $N_{s} = N_{l} =50$-th order, 
\begin{equation}
C_{5s}^{(50)}(g) = \sum_{k=0}^{50}s_{k} g^{k}, \qquad 
C_{5l}^{(50)}(g) = g^{-4}\sum_{k=0}^{50}l_{k} g^{-k}.
\label{Eq:L5-expand}
\end{equation}
Their leading order terms are 
\begin{align}
C_{5s}^{(50)}(g)
=&
2+\frac{5 g^2}{2}-\frac{3 g^3}{2}+\frac{17 g^4}{8} 
+\mathcal{O}\left(g^5\right) ,\NN\\
C_{5l}^{(50)}(g)
=&g^{-4}\left( 64-256g^{-1} +928g^{-2} -3008g^{-3} +9440g^{-4} 
+\mathcal{O}(g^{-5} )  \right) .
\end{align}
By the extrapolation in \cite{Honda:2014bza}, we can read
\begin{\eq}
N_s^\ast =50 ,\quad g_s^\ast =0.4,\quad N_l^\ast =50,\quad g_l^\ast =3.8,
\end{\eq}
and it turned out that both the expansions are convergent. %also in this case.
As shown in Figure~\ref{fig:specific}, 
the sharp peak locates outside the reliable domain, 
$g_{s}^{\ast} < g < g_{l}^{\ast}$.
Because of $N_{s}^{\ast} = N_{s}, N_{l}^{\ast} = N_{l}$, 
we can use \eqref{Eq:L5-expand} directly 
%without performing the optimal truncation 
to make the 
interpolating functions.
Several interpolating functions %for the function $C_{5}(g)$
have been given in eq.~(B.3) of \cite{Honda:2014bza} already. 
%and they are listed in \eqref{Eq:L5-Interpolate} in Appendix~\ref{app:L5}.
As in \eqref{Random-linear} and \eqref{Eq:Int-func-conv-1},
we consider 
randomly generated 
linear combinations of functions 
in eq.~(B.3) of \cite{Honda:2014bza}
%\eqref{Eq:L5-Interpolate} 
as the samples of interpolating functions.
%$c_{i}^{[r,s]}$,
%\begin{align}
%\hat{G}^{[r,s]}(g) 
%=& c_{1}^{[r,s]}C^{(-4)}_{1,1}(g)
%+c_{2}^{[r,s]}C^{(-4/3)}_{1,1}(g)
%+c_{3}^{[r,s]}C^{(-2)}_{1,2}(g)
%+c_{4}^{[r,s]}C^{(-1)}_{1,2}(g)
%\NN \\
%&+c_{5}^{[r,s]}C^{(-4)}_{2,2}(g)
%+c_{6}^{[r,s]}C^{(-2)}_{2,3}(g)
%+c_{7}^{[r,s]}C^{(-1)}_{2,3}(g)
%+c_{8}^{[r,s]}C^{(-2)}_{3,2}(g)
%\NN \\
%&
%+c_{9}^{[r,s]}C^{(-4)}_{3,3}(g)
%+c_{10}^{[r,s]}C^{(-4/7)}_{3,3}(g)
%+c_{11}^{[r,s]}C^{(-2)}_{3,4}(g)
%+c_{12}^{[r,s]}C^{(-1)}_{3,4}(g)
%\NN \\
%&+c_{13}^{[r,s]}C^{(-1)}_{4,3}(g)
%+c_{14}^{[r,s]}C^{(-4/9)}_{4,4}(g)
%+c_{15}^{[r,s]}C^{(-1)}_{4,5}(g)
%+c_{16}^{[r,s]}C^{(-1)}_{6,5}(g)
%\NN \\
%&+c_{17}^{[r,s]}C^{(-1)}_{6,7}(g)
%+c_{18}^{[r,s]}C^{(-1)}_{7,6}(g)
%\label{Eq:Int-func-L5}
%\end{align}
%where 
%\begin{equation}
%\sum_{i=1}^{18}c_{i}^{[r,s]} =  1, \qquad c_{i}^{[r,s]}\ge 0.
%\label{constraint-L5}
%\end{equation}
By using the samples, 
we calculate 
the correlation coefficients. 
%$Cr_{1,2}$ and $De_{1,2}$. %and the one between $Cr_{2}$ and $De_{2}$.
The correlation coefficients are
\begin{align}
 \rho^{[1]}_{Cr_1De_1} = 0.8031
< \rho^{[1]}_{Cr_oDe_o} = 0.88367308
< \rho^{[1]}_{Cr_2De_2} = 0.97453243,
\NN \\
 \rho^{[2]}_{Cr_1De_1} = 0.632352
<\rho^{[2]}_{Cr_oDe_o} = 0.68630313
<\rho^{[2]}_{Cr_2De_2} = 0.90483010,
\NN \\
 \rho^{[3]}_{Cr_1De_1} = 0.695725
<\rho^{[3]}_{Cr_oDe_o} = 0.78550976
<\rho^{[3]}_{Cr_2De_2} = 0.93317359,
\NN \\
 \rho^{[4]}_{Cr_1De_1} = 0.609671
<\rho^{[4]}_{Cr_oDe_o} = 0.74064316
<\rho^{[4]}_{Cr_2De_2} = 0.93436356,
\NN \\
 \rho^{[5]}_{Cr_1De_1} = 0.666964
<\rho^{[5]}_{Cr_oDe_o} = 0.76862717
<\rho^{[5]}_{Cr_2De_2} = 0.93445856.
% _3
\label{FL5rho12}
\end{align}
We should note that the correlation coefficients 
$\rho_{Cr_1De_1}$ and $\rho_{Cr_oDe_o}$ 
%between 
%$Cr_{1,o}$ and $De_{1,o}$ 
become weak,
while the correlations between $Cr_{2}$ and $De_{2}$
are still strong (bigger than $0.9$).
So the reference quantities $Cr_{1,o}$ become useless
by the appearance of the sharp peak outside the reliable domain,
while $Cr_{2}$ is still useful.
We can compare between them 
by using the plots Figure~\ref{fig:L5Cr1} also.
%(Plotype are basically same as the one in $L=2$.)

\begin{figure}[bht]
\begin{center}
\includegraphics[width=7.0cm]{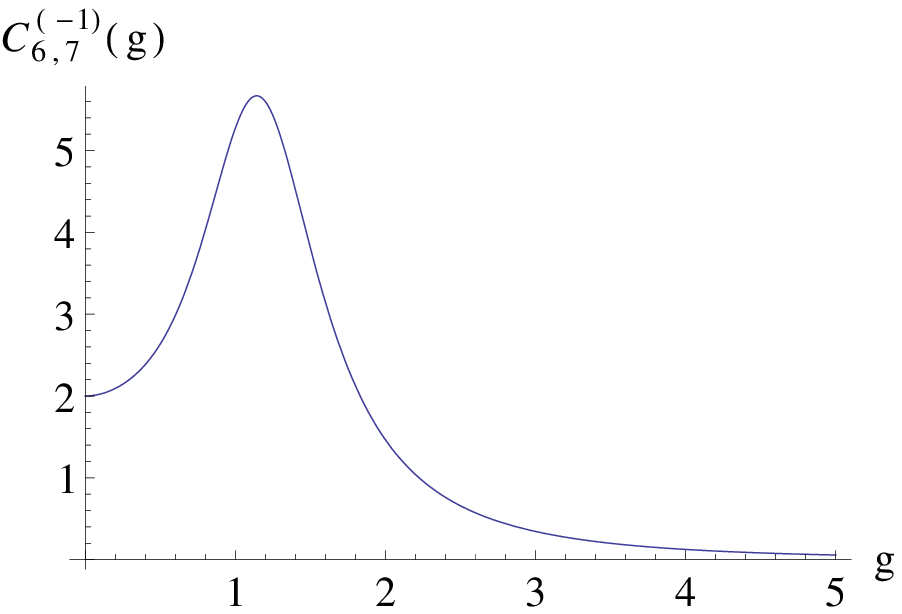}
\includegraphics[width=7.0cm]{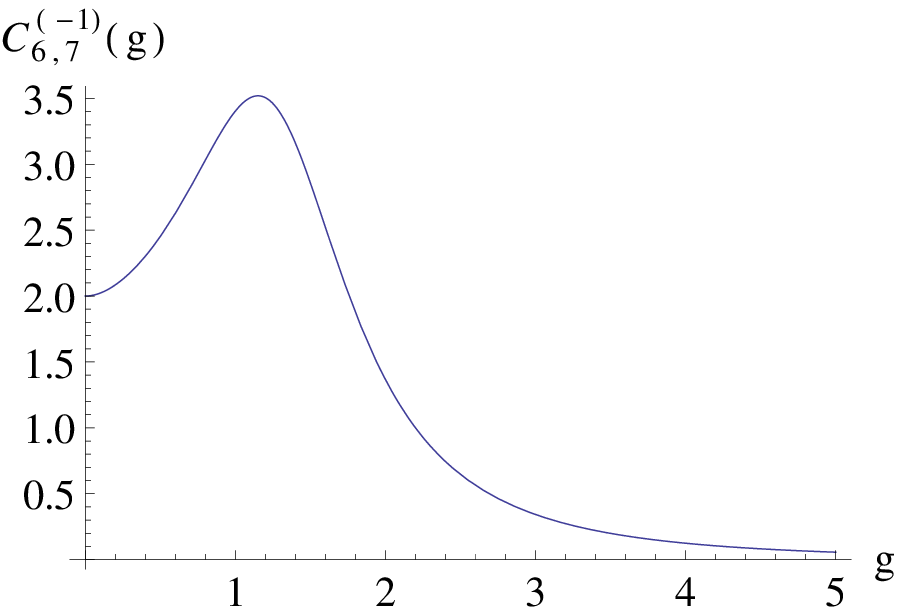}
\end{center}
\caption{[Left] Plot of $C_{6,7}^{(-1)}(g)$ to $g$ in the $L=5$ case. 
[Right] Plot of $C_{6,7}^{(-1)}(g)$ with respect to $g$ in the $L=8$ case. 
We can see that interpolating functions also have sharp peaks.}
\label{fig:peak-interpolate}
\end{figure}

One might wonder how we can notice the presence of the sharp peak 
without knowing the true function. 
Even without knowing the true function, by plotting an 
interpolating function, we can guess the presence of the sharp peak.
If a true function has a sharp peak, 
its interpolating function %$C_{m,n}^{(\alpha)}$
%with large numbers of $(m,n)$ 
tends to have a sharp peak
as shown in Figure~\ref{fig:peak-interpolate}.
This study instructs that if there is a sharp peak in an 
interpolating function, we should start to use 
$Cr_{2}$ only. %instead of $Cr_{1}$.

\subsubsection{$8 \times 8$ lattice}
\label{Sec:L8-1}
%As in case of the $L =5$, 
The function 
$C_{8}(g)$ 
for $L=8$ 
also has a sharp peak
as shown in Figure~\ref{fig:specific}.
Even if we do not know about the 
true function $C_8(g)$, 
we can deduce the presence of the peak by plotting an interpolating function 
as shown in Figure~\ref{fig:peak-interpolate}.

Also in this example, 
we assume that we know 
only the small-$g$ and large-$g$ expansions 
upto $N_{s} = N_{l} =50$-th order, 
\begin{equation}
C_{8s}^{(50)}(g) = \sum_{k=0}^{50}s_{k} g^{k}, \qquad 
C_{8l}^{(50)}(g) = g^{-4}\sum_{k=0}^{50}l_{k} g^{-k},
\label{Eq:L8-expand}
\end{equation}
where their leading order terms are
\begin{align}
C_{8s}^{(50)}(g)
=&2+\frac{5 g^2}{2}-\frac{5 g^3}{2}+\frac{29 g^4}{8} +\mathcal{O}(g^5 ) ,\NN\\
C_{8l}^{(50)}(g)
=&g^{-4}\left( 64-256g^{-1} +928g^{-2} -3008g^{-3} +9440g^{-4} 
+\mathcal{O}(g^{-5} )  \right) ,
\end{align}
respectively.
By the study in \cite{Honda:2014bza}, 
it has been known that
\begin{\eq}
N_s^\ast =50 ,\quad g_s^\ast =0.4,\quad N_l^\ast =50,\quad g_l^\ast =3.7,
\end{\eq}
and that both the expansions are convergent.
We can see that the peak locates outside the reliable domain 
as shown by Figure~\ref{fig:specific}.
As in the previous case, 
because of $N_{s}^{\ast} = N_{s}, N_{l}^{\ast} = N_{l}$, 
we can use \eqref{Eq:L8-expand} directly to make the 
interpolating functions.
Several interpolating
functions were already given in eq.~(B.4) of \cite{Honda:2014bza}.
%\eqref{Eq:L8-Interpolate}.
As in \eqref{Random-linear} and \eqref{Eq:Int-func-conv-1},
we generate interpolating functions
randomly as the samples 
by linear combinations of the functions 
in eq.~(B.4) of \cite{Honda:2014bza}.
%\eqref{Eq:L8-Interpolate}.
%with randomly generated coefficients.
%$c_{i}^{[r,s]}$,
%\begin{align}
%\hat{G}^{[r,s]}(g) 
%=& c_{1}^{[r,s]}C^{(-4)}_{1,1}(g)
%+c_{2}^{[r,s]}C^{(-4/3)}_{1,1}(g)
%+c_{3}^{[r,s]}C^{(-2)}_{1,2}(g)
%+c_{4}^{[r,s]}C^{(-1)}_{1,2}(g)
%\NN \\
%&+c_{5}^{[r,s]}C^{(-4)}_{2,2}(g)
%+c_{6}^{[r,s]}C^{(-2)}_{2,3}(g)
%+c_{7}^{[r,s]}C^{(-1)}_{2,3}(g)
%+c_{8}^{[r,s]}C^{(-1)}_{3,2}(g)
%\NN \\
%&
%+c_{9}^{[r,s]}C^{(-4)}_{3,3}(g)
%+c_{10}^{[r,s]}C^{(-4/7)}_{3,3}(g)
%+c_{11}^{[r,s]}C^{(-1)}_{3,4}(g)
%+c_{12}^{[r,s]}C^{(-1)}_{4,3}(g)
%\NN \\
%&+c_{13}^{[r,s]}C^{(-4/9)}_{4,4}(g)
%+c_{14}^{[r,s]}C^{(-1)}_{4,5}(g)
%+c_{15}^{[r,s]}C^{(-1)}_{5,6}(g)
%+c_{16}^{[r,s]}C^{(-1)}_{6,7}(g)
%\NN \\
%&+c_{17}^{[r,s]}C^{(-1)}_{7,8}(g)
%+c_{18}^{[r,s]}C^{(-1)}_{8,9}(g)
%+c_{19}^{[r,s]}C^{(-1)}_{9,10}(g)
%+c_{20}^{[r,s]}C^{(-1)}_{7,6}(g)
%\NN \\
%&+c_{21}^{[r,s]}C^{(-1)}_{8,7}(g)
%+c_{22}^{[r,s]}C^{(-1)}_{9,8}(g)
%\label{Eq:Int-func-L8}
%\end{align}
%where 
%\begin{equation}
%\sum_{i=1}^{22}c_{i}^{[r,s]} =  1, \qquad c_{i}^{[r,s]}\ge 0.
%\label{constraint-L8}
%\end{equation}
We compute the correlation coefficients 
%between $Cr_{1,2,o}$ and $De_{1,2,o}$
by using the samples.
The correlation coefficients are 
\begin{align}
 \rho^{[1]}_{Cr_1De_1} = 0.611783
< \rho^{[1]}_{Cr_oDe_o} =0.74837673
< \rho^{[1]}_{Cr_2De_2} =0.86202835,
\NN \\
 \rho^{[2]}_{Cr_1De_1} = 0.732422
<\rho^{[2]}_{Cr_oDe_o} =0.89045374
<\rho^{[2]}_{Cr_2De_2} = 0.95196590,
\NN \\
 \rho^{[3]}_{Cr_1De_1} = 0.743958
<\rho^{[3]}_{Cr_oDe_o} = 0.87655349
<\rho^{[3]}_{Cr_2De_2} = 0.95651297,
\NN \\
 \rho^{[4]}_{Cr_1De_1} = 0.66496
<\rho^{[4]}_{Cr_oDe_o} = 0.85658632
<\rho^{[4]}_{Cr_2De_2} = 0.95161652,
\NN \\
 \rho^{[5]}_{Cr_1De_1} = 0.677466
<\rho^{[5]}_{Cr_oDe_o} = 0.85909180
<\rho^{[5]}_{Cr_2De_2} = 0.94872463.
% _3
\label{FL8rho12}
\end{align}
Also in the $L=8$ case, 
$\rho_{Cr_1De_1}$ and $\rho_{Cr_oDe_o}$ become weak 
by the presence of sharp peak. On the other hand the 
correlations between $Cr_{2}$ and $De_{2}$ are 
still strong, bigger than $0.9$.
%We can compare them also by the plot in Figure~\ref{fig:L8Cr1}. 
%\begin{figure}[tb]
%\begin{center}
%\includegraphics[width=7.0cm]{FL8-Cr1De1-5.eps}
%\includegraphics[width=7.0cm]{FL8-Cr2De2-5.eps}
%\end{center}
%\caption{In the $L=8$ Ising model,
%[Left] $(Cr_1,De_1)$ obtained at the fifth examination with 100 samples
%[Right] $(Cr_2,De_2)$ at the fifth examination with 100 samples.
%We can see that 
%$(Cr_{2}, De_{2})$ have much stronger correlation while
%plot of the $(Cr_{1}, De_{1})$ are scattered.}
%\label{fig:L8Cr1}
%\end{figure}

\subsubsection{Infinite lattice}
%Finally, let us consider the case with infinite lattice $L = \infty$.
In the $L = \infty$ case, 
the true function has singularity
and there is the phase transition.
$Cr_{1}$ is obviously no longer useful in this case, 
then we will focus only on 
whether $Cr_{2,o}$ can be good reference quantities or not
even in the presence of the singularity.
\footnote{Obviously $De_{1} = \infty$
by the presence of singularity in $L = \infty$, so there is no point to 
discuss the correlation between $Cr_{1}$ and $De_{1}$.
While $De_{o}$ is still finite because the integration over $g$
around the singular point $g = \sqrt{2}$ 
\begin{equation}
\int^{\sqrt{2} + \epsilon}_{\sqrt{2}-\epsilon} dg \, C_{\infty} (g)
\sim \int^{1}_{1-\epsilon'}\frac{dx }{\sqrt{1-x}} K(x) < \infty
\end{equation}
is not infinite.}

The true function for the specific heat in the $L = \infty$ case is given by
\begin{\eqa}
C_\infty (g)
&=& \frac{16 (g+1)}{\pi  g (g+2)} 
\Biggl[ K\left(\frac{4 g (g+1) (g+2)}{\left(g^2+2 g+2\right)^2}\right) 
-E\left( \frac{4 g (g+1) (g+2)}{\left(g^2+2 g+2\right)^2}\right) \NN\\
&& -\left( \frac{2 (g+1)}{ (g+1)^2+1} \right)^2
 \left\{ \frac{g^4+4g^3-8 g-4}{\left(g^2+2 g+2\right)^2} K\left( \frac{4 g (g+1) (g+2)}{\left(g^2+2 g+2\right)^2}\right) +\frac{\pi}{2} \right\} 
\Biggr] ,
\end{\eqa}
where $K(z)$ and $E(z)$ are 
$$K(z) = \int_0^{\pi /2} dt ( 1-z\sin^2{t} )^{-1/2},$$
$$E(z) = \int_0^{\pi /2} dt ( 1-z\sin^2{t} )^{1/2}.$$
%the complete elliptic integrals of the first and second kinds
%respectively.
%\footnote{
%These are defined by
%} 
As in the previous cases, 
we will assume that we know only 
the small-$g$ and large-$g$ expansions 
upto $N_{s} = N_{l} =50$-th order, 
\begin{equation}
C_{\infty s}^{(50)}(g) = \sum_{k=0}^{50}s_{k} g^{k}, \qquad 
C_{\infty l}^{(50)}(g) = g^{-2}\sum_{k=0}^{50}l_{k} g^{-k},
\label{Eq:Linf-expand}
\end{equation}
their leading order terms are
\begin{align}
C_{\infty s}^{(50)}(g)
=&6-\frac{11 g}{4}-\frac{15 g^2}{4}+\frac{3265 g^3}{256}-\frac{651 g^4}{64}
 +\mathcal{O}(g^5 ) ,\NN\\
C_{\infty l}^{(50)}(g)
=&g^{-2}\left( 16 -72g^{-1} +164g^{-2} +15g^{-3}  -\frac{3087}{4}g^{-4} 
+\mathcal{O}(g^{-5} )  \right) ,
\label{Eq:Linf-expand-lead}
\end{align}
respectively.
%As in the finite lattice cases,
By the extrapolation in \cite{Honda:2014bza}, 
we obtain
\begin{\eq}
N_s^\ast =50 ,\quad g_s^\ast =1,\quad N_l^\ast =50,\quad g_l^\ast =2,
\end{\eq}
and we can see that both the expansions are convergent.
So we can use \eqref{Eq:Linf-expand} directly to make the 
interpolating functions.
Several interpolating functions are already given by 
eq.~(B.5) of \cite{Honda:2014bza}. %listed in \eqref{Eq:Linf-Interpolate}.
We generate interpolating functions randomly
by the linear combinations of functions 
in eq.~(B.5) of \cite{Honda:2014bza}
%\eqref{Eq:Linf-Interpolate}
as in the previous cases.
%interpolating functions in 
% with randomly chosen coefficients
%$c_{i}^{[r,s]}$,
%\begin{align}
%\hat{G}^{[r,s]}(g) 
%=& c_{1}^{[r,s]}C^{(-2/3)}_{1,1}(g)
%+c_{2}^{[r,s]}C^{(-1)}_{2,1}(g)
%+c_{3}^{[r,s]}C^{(-2/5)}_{2,2}(g)
%+c_{4}^{[r,s]}C^{(-1)}_{2,3}(g)
%\NN \\
%&+c_{5}^{[r,s]}C^{(-1)}_{3,4}(g)
%+c_{6}^{[r,s]}C^{(-2/9)}_{4,4}(g)
%+c_{7}^{[r,s]}C^{(-1)}_{4,5}(g)
%+c_{8}^{[r,s]}C^{(-2/11)}_{5,5}(g)
%\NN \\
%&
%+c_{9}^{[r,s]}C^{(-1)}_{5,6}(g)
%+c_{10}^{[r,s]}C^{(-2/13)}_{6,6}(g)
%+c_{11}^{[r,s]}C^{(-1)}_{6,7}(g)
%+c_{12}^{[r,s]}C^{(-1)}_{7,6}(g)
%\NN \\
%&+c_{13}^{[r,s]}C^{(-2/15)}_{7,7}(g)
%+c_{14}^{[r,s]}C^{(-1)}_{7,8}(g)
%+c_{15}^{[r,s]}C^{(-1)}_{9,8}(g)
%+c_{16}^{[r,s]}C^{(-1)}_{9,10}(g)
%\NN \\
%&+c_{17}^{[r,s]}C^{(-1)}_{10,9}(g)
%\label{Eq:Int-func-Linf}
%\end{align}
%where 
%\begin{equation}
%\sum_{i=1}^{17}c_{i}^{[r,s]} =  1, \qquad c_{i}^{[r,s]}\ge 0.
%\label{constraint-Linf}
%\end{equation}
By using the generated functions as the samples,
we check the reference quanities 
$Cr_{2,o}$ by calculating the correlation coefficients. 
The correlation coefficients between $Cr_{2}$ and $De_{2}$ are
\begin{align}
\rho^{[1]}_{Cr_2De_2} =0.9889143983,
\quad 
\rho^{[2]}_{Cr_2De_2} = 0.9857282660,
\quad
\rho^{[3]}_{Cr_2De_2} = 0.9746687539,
\NN \\
\rho^{[4]}_{Cr_2De_2} = 0.9835468034,
\quad
\rho^{[5]}_{Cr_2De_2} = 0.9745817796.
\label{FLinf-rho2}
\end{align}
Even in the presence of the 
singularity, $Cr_{2}$ has strong 
correlation with $De_{2}$. Hence
the $Cr_{2}$ can be a reliable reference.

\begin{figure}[tbp]
\begin{center}
\includegraphics[width=7.0cm]{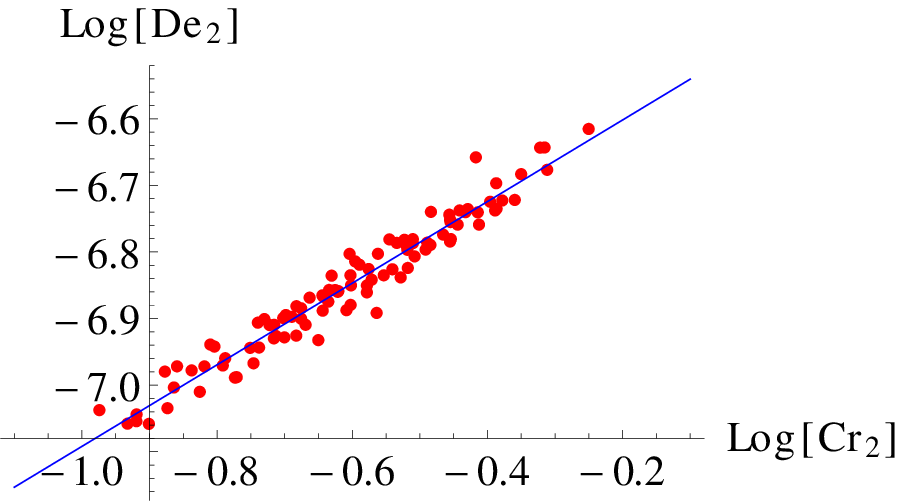}
\includegraphics[width=7.0cm]{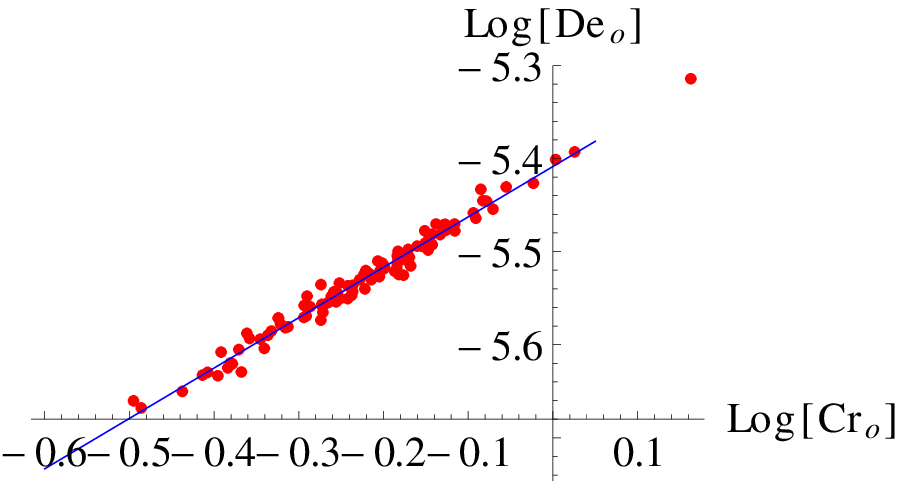}
\end{center}
\caption{Plots of $(Cr_2,De_2)$ and 
$(Cr_o,De_o)$
during the fifth calculation of the correlation coefficient
in case of the $L = \infty$ Ising model.}
\label{fig:LinfCr2Cro}
\end{figure}

%Actually, in $L = \infty$ case, 
The correlation coefficients between 
$Cr_{o}$ and $De_{o}$ are also strong, 
\begin{align}
\rho^{[1]}_{Cr_oDe_o} =0.970161493,
\quad 
\rho^{[2]}_{Cr_oDe_o} = 0.973184305,
\quad
\rho^{[3]}_{Cr_oDe_o} = 0.986896467,
\NN \\
\rho^{[4]}_{Cr_oDe_o} = 0.9928459634,
\quad
\rho^{[5]}_{Cr_oDe_o} = 0.9903143989.
% _2
\label{FLinf-rhoo}
\end{align}
They are even stronger than the ones between $Cr_{2}$ and $De_{2}$.
But unlike the $L = 5,8$ cases, 
$\rho_{Cr_2De_2}$ are also 
strong enough and there are not so much differences between 
$\rho_{Cr_2De_2}$ and $\rho_{Cr_oDe_o}$. 
Also from the plots in Fig.~\ref{fig:LinfCr2Cro}, 
we can see that there are not so much differences 
between $\rho_{Cr_2De_2}$ and $\rho_{Cr_oDe_o}$.
%The plots of the $(Cr_2, De_2)$ are put in 
%Figure~\ref{fig:LinfCr2}.
%\subsection{Summary of this section}

\section{Conclusion $\&$ discussion} 
\label{Conclusion}
%{\bf To be editted properly.}
%For establishing the prescription
%to choose a good interpolating functions,
We suggested the %a set of 
quantities $Cr_{1,2,o}$
given in \eqref{CR12}, \eqref{CR22} and \eqref{CRo2}
as reference quantities for choosing 
a good interpolating function.
To check whether they are reliable reference quantities
or not,
we calculated correlation coefficients
between $Cr_{j}$ and $De_{j}$ for $j \in \{1,2,o\}$ %
%$Cr_{1}$ and $De_{1}$ (denoted by $\rho^{[i]}_{Cr_1De_1}$), 
%the ones between $Cr_{2}$ and $De_{2}$ (denoted by $\rho^{[i]}_{Cr_2De_2}$),
%and the ones between $Cr_{o}$ and $De_{o}$
%(denoted by $\rho^{[i]}_{Cr_oDe_o}$)
for several examples.
Here $De_{1,2,o}$ are given by \eqref{DE1}, \eqref{DE2} 
and \eqref{DEo}. 
%while the correlation coefficients are
%denoted by $\rho_{Cr_{1}De_{1}}$, 
%$\rho_{Cr_{2}De_{2}}$ and $\rho_{Cr_{o}De_{o}}$. 
%, they indicate the actual degree of deviation
%between interpolating function and deviation. 
%In addition to these correlation coefficients, 
%By the examination, 
We observed
\begin{enumerate}
 \item 
For functions without sharp peak
outside the reliable domain, 
all $Cr_{1,2,o}$ can be good reference quantities for 
choosing a good interpolating function.
Among them $Cr_{1}$ seems to be the best reference quantity. 
This comes from the observations in  \eqref{F11rho1}$\sim$\eqref{F11rhoo},
\eqref{Fphirho1}$\sim$\eqref{Fphirhoo},
%\eqref{Cr1De1-lattice}$\sim$\eqref{Fphirhoo} and 
\eqref{Cr1De1-lattice}$\sim$\eqref{Cr2De2-lattice} and 
\eqref{FBasymrho1}$\sim$\eqref{FBasymrhoo}.
%the correlation coefficients between $Cr_{1}$ and $De_{1}$
%are generically stronger than the ones between $Cr_{2}$ and $De_{2}$
%in the examples where the functions does not have sharp peak
%outside the reliable domain.
%This means that $Cr_{1}$ will be better reference quantity than $Cr_2$
%for the true function without sharp peak outside the reliable domain.
\item 
For functions with sharp peak outside the 
reliable domain, only $Cr_{2}$ can be reliable.
This comes from the observations in \eqref{FL5rho12}, \eqref{FL8rho12} and
\eqref{FLinf-rho2}.
%and \eqref{FLinf-rhoo}.
\end{enumerate}
%As a summary we can say that
%$Cr_{1}$ is a powerful criteria to select a good 
%interpolating function if there is not the sharp peak in the 
%intermediate region which is outside of the reliable domain.
Usually %the presence of 
the sharp peak in the true function
can be found by plotting the interpolating functions.
This observation indicates that
the combination of $Cr_{1}$ and $Cr_{2}$ works 
as the best reference for selecting a good interpolating function. 
It is also convenient to use $Cr_{2}$ only, because
only $Cr_{2}$ has universal usage independent of functions.

%In order to make the criterion in 
%\cite{Honda:2014bza} more sufficient, 
%We confirmed that 
%the reference quantity $Cr_{o}$ works well 
%to deduce the $De_2$ to some extent.
%We have also compared 
%the correlation coefficients with the ones 
%between 
%$Cr_{o}$ with the set of the quantities
%$Cr_{1}$ and $Cr_2$ suggested in this paper,
%By the comparison, 
%So combining above two observation, 
%from the above two observation, 
%our suggesting set of 
%The above observations 1., 2., A., and B. indicate that
%the prescription based on 
%observation 
%quantities $Cr_{1,2}$ work better than $Cr_{o}$
%as reference quantities for selecting a good interpolating function.

The analysis in this paper requires the large order expansions
which was true for the examples we considered.
But in many cases, we may not have 
large order expansions.
So, it is important to establish a way to estimate
$g_{s}^{\ast}, g_{l}^{\ast}, N_{s}^{\ast}$ and $N_{l}^{\ast}$
also for the cases 
where we have limited number of expansions.

\subsection*{Acknowledgment}
We are grateful to Masazumi Honda and Ashoke Sen for their collaboration 
at the early stage of this project.
Especially, we would like to thank Ashoke Sen for suggesting 
the author to write down this paper and kindly 
reading the manuscript carefully.
We thank Abhishek Chowdhury, 
Yoshinori Honma, Swapnamay Mondal, 
Satchitananda Naik, 
Kenji Nishiwaki,  Kenji Ogawa, Tetsuya Onogi, Roji Pius and  Koji Tsumura
for valuable discussions and comments.
In particular, we acknowledge Roji Pius 
for reading the manuscript.
T.T would like to express the gratitude for the hospitality
offered by 
NTU (particularly Pei-Ming Ho and Heng-Yu Chen and related secretaries), 
Weizmann institute (particularly Ofer Aharony and related secretaries), ICTP
(particularly Kumar Narain and related secretaries)
during his stay.
He is also thankful to the kind support provided 
by the Strings 2014 organizers.
Finally he wishes to acknowledge the kind support provided by the Indian people.
\appendix
\section{Insufficiency of 
the criterion given in \cite{Honda:2014bza}}
\label{Sec:Honda-judge}
As discussed in subsection \ref{sec:landscape}, 
there are uncountably infinite number of interpolating functions,
because a linear combination of interpolating functions 
is also an interpolating function again.
So it is important to establish a criterion
for selecting a good interpolating function.
In \cite{Honda:2014bza}, the authors proposed such a criterion,
but we will argue that their criterion is insufficient.

Let us start by a brief 
review of the analysis given in \cite{Honda:2014bza}. 
First of all, they picked up one subset from 
infinite number of interpolating functions $G(g)$ 
%corresponding to each ``true function'' $F(g)$, 
%of physical quantities, 
where $g$ is the parameter. 
We should note that 
this subset has no %was chosen casually, there is no
particular significance 
compared to any %on the subset, 
other possible subsets. 
%should be dealt with equivalently.
Within the selected subset they observed the following matching 
%with respect to the test interpolating function
%$G(g)$,
\begin{align}
&\text{ (the function $G(g)$ with the smallest 
$I_s[G(g)] + I_l[G(g)]$ 
in the subset)}
\NN \\
=&\text{(the function 
$G(g)$
with the smallest $\Lambda^{-1}\int_{0}^{\Lambda} dg 
\left|\frac{G(g)- F(g)}{F(g)}\right|$
 in the subset)},
\label{statement}
\end{align}
here $\Lambda^{-1}\int_{0}^{\Lambda} dg 
\left|\frac{G(g)- F(g)}{F(g)}\right|$
measures deviation 
between $G(g)$ and the true function $F(g)$. 
$I_s[G(g)], I_{l}[G(g)]$ are defined 
in (3.1) of \cite{Honda:2014bza},
\begin{equation}
I_{s}[G] =\int^{g^{\ast}_{s}}_{0} dg \left|
G(g) - F^{(N_s^{\ast})}_{s}(g)
\right|, \quad 
I_{l}[G] =\int_{g^{\ast}_{l}}^{\Lambda} dg \left|
G(g) - F^{(N_l^{\ast})}_{l}(g)
\right|
\end{equation}
where $F_{s}^{(N_s^{\ast})}$ and $F_{l}^{(N_l^{\ast})}$
are the small-$g$ and large-$g$ expansions up to order
$N_{s}^{\ast}$ and $N_{l}^{\ast}$ respectively.
Here the validity of the 
expansions $F_{s}^{(N_s^{\ast})}$ and $F_{l}^{(N_l^{\ast})}$
are limited to the domains $0 < g < g_{s}^{\ast}$ and $g_{l}^{\ast} < g$ 
respectively.
(In our discussions, we denoted $I_{s}+ I_{l}$ by $Cr$ and
$\Lambda^{-1}\int_{0}^{\Lambda} dg \left|\frac{G(g)- F(g)}{F(g)}\right|$
by $De$.)
Based on the above observation 
\eqref{statement} which is true within the selected subset,
they asserted that 
the best interpolating function has 
the minimum value for $I_s[G]+I_{l}[G]$.
But we should remember that the notion of the criterion should be 
independent of the choice of subset.
By constructing an explicit example, 
we will show in the following subsection that 
it is possible to choose another subset
where the observation \eqref{statement} is not valid.
This means that their criterion in \cite{Honda:2014bza}
is insufficient as a universal criterion, which 
should be independent of the choice of the subset.
%This means %the above assertion 
%the best interpolating function which 
%has the smallest deviation from the true function 
%may not have the least value for $I_s[G] + I_{l}[G]$.

\subsection{A counter example}
Let us re-examine the case 
of the two-dimensional Ising model with lattice size $L = 2$
considered in subsubsection 4.2.1 of \cite{Honda:2014bza}. 
There they tried to find interpolating function 
for the specific heat $C_{2}(g)$.
Here the parameter $g = e^{2J/T} -1$ where $J$ is the 
coupling constant of Ising model and $T$ is the temperature.
They computed $De = \Lambda^{-1}\int dg 
\left|\frac{C_{m,n}^{\alpha}- C_2}{C_2}\right|$
and $Cr = I_{s} + I_{l}$ 
for a class of interpolating functions $C_{m,n}^{(\alpha)}$, 
which are listed in Table~\ref{tab:L2}.
For this set of interpolating functions given in the 
Table~\ref{tab:L2}, 
the observation \eqref{statement} is true.

\begin{table}[h]
\begin{center}
  \begin{tabular}{|c|c|c|c|c|}
  \hline & $\Lambda^{-1}\int dg \bigl| \frac{C_{m,n}^{(\alpha )}-C_2}{C_2} \bigr|$ &$I_s [C_{m,n}^{(\alpha )}]$ &$I_l [C_{m,n}^{(\alpha )}]$ & $I_s +I_l$\\
  \hline\hline $C_{1,1}^{(-4)} $   &0.00224809 &0.0912750 &0.102638 &0.193913 \\
\hline
$C_{1,1}^{(-4/3)} $   &0.000817041 &0.0617656 &0.0193989 &0.0811645 \\
\hline
 $C_{1,2}^{(-2)} $   &0.00100228 &0.0751219 &0.0466568 & 0.121779\\
\hline
 $C_{1,2}^{(-1)} $   &0.000286070 &0.058021  &0.00257514 &0.0605961 \\
\hline
 $C_{2,2}^{(-4)} $   &0.000173889  &0.00768450   &0.00852503 &0.0162095 \\
\hline
 $C_{2,3}^{(-2)} $   &0.000158806 &0.00224879  &0.00550386 &0.00775265\\
\hline
 $C_{3,2}^{(-2)} $   &0.000322997  &0.00658097  &0.0116865 &0.0182674\\
\hline
 $C_{3,4}^{(-1)} $   &0.0000147709  &0.000148814  &0.000258785 &0.000407598\\
\hline
 $C_{4,3}^{(-2)} $   &0.000168121  &0.00345741 &0.00565649 &0.0091139\\
\hline
 $C_{4,3}^{(-1)} $   &0.0000651441 &0.00156065  &0.00170668 &0.00326733\\
\hline
 $C_{5,4}^{(-1)} $   &0.0000207392  &0.000525855  &0.000327812 &0.000853666\\
\hline
 $C_{6,5}^{(-1)} $   &0.0000119340  &0.000174690 &0.000192164 &0.000366854\\
\hline
 $\bf C_{7,6}^{(-1)} $   &$\bf 1.22853\times 10^{-6}$  &$\bf 3.39663\times 10^{-6}$ 
&$\bf 0.0000523107$ &\bf 0.0000557073\\
\hline
 $C_{6,7}^{(-1)} $   &0.0000128648  &0.000129274  &0.0000598797 &0.000189154\\
\hline
  \end{tabular}
\caption{$Cr$  and $De$ in the original subset in \cite{Honda:2014bza}}
\label{tab:L2}
\end{center}
\end{table}
\begin{table}[h]
\begin{center}
  \begin{tabular}{|c|c|c|c|c|}
  \hline & $\Lambda^{-1}\int dg \bigl| \frac{C_{m,n}^{(\alpha )}-C_2}{C_2} \bigr|$ &$I_s [C_{m,n}^{(\alpha )}]$ &$I_l [C_{m,n}^{(\alpha )}]$ & $I_s +I_l$\\
  \hline\hline $C_{1,1}^{(-4)} $   &0.00224809 &0.0912750 &0.102638 &0.193913 \\
\hline
$C_{1,1}^{(-4/3)} $   &0.000817041 &0.0617656 &0.0193989 &0.0811645 \\
\hline
 $C_{1,2}^{(-2)} $   &0.00100228 &0.0751219 &0.0466568 & 0.121779\\
\hline
 $C_{1,2}^{(-1)} $   &0.000286070 &0.058021  &0.00257514 &0.0605961 \\
\hline
 $C_{2,2}^{(-4)} $   &0.000173889  &0.00768450   &0.00852503 &0.0162095 \\
\hline
 $C_{2,3}^{(-2)} $   &0.000158806 &0.00224879  &0.00550386 &0.00775265\\
\hline
 $C_{3,2}^{(-2)} $   &0.000322997  &0.00658097  &0.0116865 &0.0182674\\
\hline
 $C_{3,4}^{(-1)} $   &0.0000147709  &0.000148814  &0.000258785 &0.000407598\\
\hline
 $C_{4,3}^{(-2)} $   &0.000168121  &0.00345741 &0.00565649 &0.0091139\\
\hline
 $C_{4,3}^{(-1)} $   &0.0000651441 &0.00156065  &0.00170668 &0.00326733\\
\hline
 $C_{5,4}^{(-1)} $   &0.0000207392  &0.000525855  &0.000327812 &0.000853666\\
\hline
 $C_{6,5}^{(-1)} $   &\bf 0.0000119340  &0.000174690 &0.000192164 &0.000366854\\
\hline
 $C_{6,7}^{(-1)} $   &0.0000128648  &0.000129274  &0.0000598797 &\bf 0.000189154\\
\hline
  \end{tabular}
\caption{$Cr$ and $De$ in a smaller subset excluding $C_{7,6}^{(-1)}$}
\label{tab:L2-small}
\end{center}
\end{table}

Now let us construct another 
Table~\ref{tab:L2-small}, 
just by deleting the second last raw of 
Table~\ref{tab:L2}.
We can view this table as made of another set of interpolating functions.
For this set of interpolating functions, 
it is not difficult to see that 
the observation \eqref{statement} is not true.
Because though minimal deviation from the actual function $C_2(g)$
happens for $C_{6,5}^{(-1)}(g)$,
the minimal $I_s+I_l$ happens for another interpolating 
function $C_{6,7}^{(-1)}(g)$.
This observation clearly invalidate the criterion provided in 
\cite{Honda:2014bza}, for choosing the best interpolating function.

%_o _2
%%%%%%%%%%%%%%%%%%%%%%%%%%%%%%%%%%%%%%%%%%%%%
%%%%%%%%%%%%%%%%%%%%%%%%%%%%%%%%%%%%%%%%%%%%%
%%%%%%%%%%%%%%%%%%%%%%%%%%%%%%%%%%%%%%%%%%%%%

\section{Reliable domain, optimal truncation by fitting.}
\label{Sec:Rel-opt}
In this Appendix, we explain
how to determine $g_{s}^{\ast}, g_{l}^{\ast}, N_{s}^{\ast}$ and $N_{l}^{\ast}$
based on the expansions $F_{s}^{(N_s)}(g)$ and $F_{l}^{(N_l)}(g)$ 
in \eqref{eq:asymptotics}.
We assume that both $N_s$ and $N_l$ are finite but large enough.

First of all, we should clarify whether 
$F_{s}^{(N_s)}(g)$ is convergent expansion or not.
In principle, if an expansion is a finite order expansion,
it is impossible to assert 
whether the expansion is asymptotic or convergent.
But if $N_s$ is large enough,
we can sometimes deduce whether the expansion is convergent or not
by extrapolating the expansion coefficients 
as in \cite{Honda:2014bza}. 
%After fitting the data, we will deduce the behavior of the coefficients
%$s_{n}$ or $l_{n}$ at large-$n$. 
By the extrapolation, if the ratio turns out to behave
\begin{equation}
\left|\frac{s_{n}}{s_{n+1}}\right| \to 0, 
 \qquad n \to \infty,  
\label{ratio}
\end{equation}
we expect that $F^{(N_s)}_{s}(g)$ is a part of an
asymptotic non-convergent expansion. While if the ratio goes to 
finite non-zero quantity, 
we think the expansion as a part of a convergent 
expansion. 
Also for large-$g$ expansion,
we apply the same way to clarify 
whether $F^{(N_l)}_l(g)$ is convergent or not.

%{\bf From here 9/25}
Depending on whether the expansion 
is convergent or not,
we apply different approaches to determine
the values $N_s^{\ast},N_l^{\ast},g^{\ast}_{s}$ and $g_l^{\ast}$.

\subsection{Asymptotic expansion}
\label{Sec:Optimize}
Let us consider the case 
that the small-$g$ expansion $F_{s}^{(N_s)}(g)$ is asymptotic.
Here we proceed to the discussion with keeping $g =g_{o}$ fixed
for a while.
%Different from the convergent expansion, 
In case of the asymptotic series, higher order expansions are not always
closer to the true function.
So the expansion truncated at the suitable order is the closest to the 
true function $F(g_{o})$ at fixed $g= g_o$. 
Such a truncation is called as the optimal truncation,
and $N_{s}^{\ast}$ represents the order at which the truncation is implemented.
In this subsection, we explain how to determine
$N_s^{\ast}$ by the fitting in case of the asymptotic expansion.

Let us consider a small-$g$ power series asymptotic expansion 
with following form
\begin{equation}
F^{(N_s)}_{s}(g) = g^{\tilde{a}}\sum_{n=0}^{{\rm floor}(\frac{N_s}{p})} a_n g^{pn},
\label{Eq:Asymp}
\end{equation}
where $p$ is a positive integer.
In large $n$, we often observe that the ratio of the 
coefficients behaves as %as linear of $n$ as
\begin{equation}
\left|
\frac{a_{n+1}}{a_{n}}
\right| \sim A n, \quad n \gg 1,
\end{equation}
where $A$ is a constant.
In this case, coefficients in large-$n$ behave as
\begin{equation}
a_{n}  \sim c n! A^{n}, \label{eq:coeffbehavior}
\end{equation}
where $c$ is a constant.
We often put the ansatz \eqref{eq:coeffbehavior}
where $c$ and $A$ will be determined
by the fitting.
If the coefficients behave as 
\eqref{eq:coeffbehavior}, we can apply the method in 
\cite{Marino:2012zq} to evaluate 
$N_{s}^{\ast}$ and $g_{s}^{\ast}$.
%in which the asymptotic expansion can be close enough
%$to the true function.
%First let us estimate the order at which the optimal truncation occur
%at fixed $g = g_{o}$.
Since the optimal truncation implemented
at the order $n = \tilde{N}_{s}^{\ast}(g_o)$ 
provides the minimum absolute value of the expansion terms,  
it requires
\begin{equation}
\frac{\partial}{\partial n}
\log |c A^{n} n! g_{o}^{pn}| \biggl|_{n = \tilde{N}^{\ast}_{s}(g_{o})} = 0.
\label{optimal}
\end{equation}
From \eqref{optimal},
by using the 
Stirling approximation $\log n! = n \left(
\log n -1\right)$,
$\tilde{N}^{\ast}_{s}(g_{o})$ is expressed in terms of 
$g_{o}$  as follows
\begin{equation}
\tilde{N}^{\ast}_{s}(g_{o}) = \left|
\frac{1}{A g_{o}^{p}}
\right|. \label{eq:N-g}
\end{equation}
%$\tilde{N}^{\ast}_{s}$ will be determined 
%if we determine the $g_{o}$, 

We consider the ``reliable domain'' next.
The ``error'' of the optimal truncation 
implemented at order $\tilde{N}_{s}^{\ast}$ 
is defined as
\begin{equation}
\delta(g;\tilde{N}^{\ast}_{s}) = |a_{\tilde{N}^{\ast}_{s}+1} 
g^{p(\tilde{N}^{\ast}_{s}+1)+\tilde{a}} |.
\end{equation}
By using this error, we define the ``reliable domain with $\epsilon$'' 
as
\begin{equation}
 0 \le g \le g_{s}^{\ast},
\end{equation}
such that
\begin{equation}
0<g < g_{s}^{\ast} 
\Rightarrow \delta(g, \tilde{N}_s^{\ast}(g_{s}^{\ast})) < \epsilon, \qquad 
\delta(g_{s}^{\ast}, \tilde{N}_s^{\ast}(g_{s}^{\ast})) = \epsilon.
\end{equation}
If once we specify the value $\epsilon$, we can 
determine $\tilde{N}^{\ast}_{s}$ 
and $g^{\ast}_{s}$ uniquely
by solving
\begin{equation}
\epsilon = \delta(g^{\ast}_{s};\tilde{N}^{\ast}_{s}(g^{\ast}_{s})), \quad 
\tilde{N}^{\ast}_{s}(g_{s}^{\ast}) = \left|
\frac{1}{A (g_{s}^{\ast})^{p}}
\right|. \label{epsilon-eq}
\end{equation}
By using $\tilde{N}_{s}^{\ast}$,
the integer $N_{s}^{\ast}$ is obtained %from $\tilde{N}_{s}^{\ast}$
as 
\begin{equation}
N_{s}^{\ast} = p \tilde{N}_{s}^{\ast}.
\end{equation}

We can 
obtain $g_{l}^{\ast}, N_{l}^{\ast}$
for an asymptotic large-$g$ expansion
by an analogous approach. 

\subsection{Convergent expansion}
In case of a convergent expansion, 
higher order expansion becomes closer 
to the true function at fixed $g$ inside
the convergent radius. 
%Hence $N_s^{\ast} = N_{l}^{\ast} = \infty$
%if we know all order of the expansion.
%{\bf From here}
Hence 
$N_{s}^{\ast} = N_{s}$ and $N_{l}^{\ast} = N_{l}$
for the convergent expansions $F^{(N_s)}_{s}(g)$ and $F^{(N_l)}_{l}(g)$
respectively.

We consider the reliable domain next.
Let us consider the small-$g$ expansion as an example.
If the convergent radius of the small-$g$
expansion is given as $\tilde{g}_{s}$, 
the expansion at $g< \tilde{g}_s$
is sufficiently close to the true function.
Hence $g_{s}^{\ast}$ should be 
$g_{s}^{\ast} = \tilde{g}_{s}$.
But because we do not know an infinite 
order expansions when we apply the 
interpolating functional methods, 
we can not know the convergent radius
in principle.
However if $N_{s}$ is large enough, 
we can deduce the convergent radius by the extrapolation.
By extrapolating the ratio, we will deduce the convergent radius as
\begin{equation}
g^{c}_{s}  \sim \left|
\frac{s_{n}}{s_{n+1}}
\right| \quad \text{at} \quad n \gg 1.
\end{equation}
%This $g_{s}^{c}$ is a deduced convergent radius obtained by fitting.
To obtain $g_{s}^{\ast}$, we need further discussion.
According to the prescription in \cite{Honda:2014bza}, 
we should also take care of the blow up point of the 
$F_{s}^{(N_s)} (g)$. The blow up point is denoted by 
$g_{s}^{b}$.
This blow up point can be found by plotting the curvature of 
$F_{s}^{(N_s)} (g)$ to $g$,
%\footnote{
because 
the blow up point locates around
the peak of the curvature.
\footnote{Since the peak has finite width, we need to 
take $g_s^{b}, g_{l}^{b}$ with avoiding the finite width.}
From these, we determine the 
supremum of the reliable domain $g_{s}^{\ast}$
for the small-$g$ expansion as
\begin{equation}
g_s^{\ast} = \min (g_{s}^c, g_{s}^b).
\end{equation}
$g_{l}^{\ast}$ in the convergent
large-$g$ expansion is also determined analogously.% obtained.
\section{Correlation between the maximum point and 
the actual degree of deviation}
%We can expect that 
If the deviation between the interpolating function 
and the true function is smaller,
the form of the interpolating function 
may be closer to the one of the 
true function. 
So one may wonder following: 
If the interpolating function is closer to the true function, 
maximum point of the interpolating function may be 
closer to the one of the true function.
%{\bf From here}
If it is true, 
the maximum point of the interpolating function 
may be useful 
for deducing the phase transition point
(or point of the sharp peak).
We will check it 
by calculating the correlation coefficients 
between 
$De_{2}$ (see \eqref{DE2}) 
and $Pd$ given by
\begin{equation}
Pd =g_{true} -g_{int}
\label{eq:distance}
\end{equation}
where 
$g_{true}$ is the maximum point of the true function 
while $g_{int}$ is the one of the interpolating 
function.
In this section, 
by using the 
functions of specific heat in the two dimensional 
Ising model with $L=5, L=8$ and $ L=\infty$ as examples, 
we check the 
correlation coefficients $\rho_{De_{2}Pd}$
between $De_{2}$ and $Pd$.
The results are listed in the Table \ref{tab:Lpeak}.
\footnote{In case of the $L = \infty$, 
to observe the maximum point of the interpolating function 
clearly, we have used the following linear combinations as samples, 
\begin{equation}
\hat{C}^{[r,s]}(g)=
a^{[r,s]} C^{(-1)}_{9,10}(g) 
+(1-a^{[r,s]}) C^{(-1)}_{10,9}(g), \qquad a^{[r,s]} \ge 0,
\end{equation}
where $a^{[r,s]}$ are randomly chosen coefficients. 
}
\begin{table}[h]
\begin{center}
  \begin{tabular}{|c|c|c|c|c|c|}
  \hline & $ \rho^{[1]}_{De_2Pd}$ &$\rho^{[2]}_{De_2Pd}$ 
&$\rho^{[3]}_{De_2Pd}$ & $\rho^{[4]}_{De_2Pd}$ & $\rho^{[5]}_{De_2Pd}$\\
  \hline\hline 
$L = 5 $   &0.790324 & 0.610306& 0.0615464& 0.0615464 & 0.193946\\
\hline
$L = 8 $   &0.575574 & 0.655114& 0.650745& 0.543775 & 0.64113\\
\hline
$L = \infty $   &0.0210254 &-0.332933 &-0.243996 &-0.145259 &-0.0465137\\
\hline
  \end{tabular}
\caption{Result of $\rho_{De_{2}Pd}$ in $L =5,8,\infty$ cases}
\label{tab:Lpeak}
\end{center}
\end{table}
From this table, it turns out that 
$Pd$ is not strongly correlated with $De_{2}$. 
\footnote{Actually, a lot of results here are not
statistically significant. So we need more samples for a firm study.}
So even if 
the deviation between the interpolating function and the true function 
is smaller,
the maximum point of the interpolating function will not always be
closer to the one of the true function.
The study in \cite{Honda:2014bza} said that
\begin{equation}
\frac{g_{true}-g_{int}}{g_{true}} = 2 \% \sim 16 \% .
\end{equation}
These differences will not be smaller even if we 
find better interpolating functions.

\section{Interpolating functions}

\subsection{$F(g) = (1-\frac{g}{5}+g^2)^{\frac{1}{2}}$}
\label{Eq:F11}
\begin{\eqa}
&&\scriptscriptstyle 
F^{(-1)}_{1,1}(g) = 
\frac{g^2+\frac{9 g}{10}+1}{g+1},
\quad
F^{(-1)}_{2,2}(g) =
\frac{g^3+\frac{27 g^2}{20}+\frac{27 g}{20}+1}{g^2+\frac{29 g}{20}+1},
\quad
F^{(-1)}_{3,3}(g)= 
\frac{g^4+\frac{9 g^3}{5}+\frac{441 g^2}{200}+\frac{9 g}{5}+1}{g^3+\frac{19
   g^2}{10}+\frac{19 g}{10}+1},
\NN\\
&&\scriptscriptstyle 
F^{(-1)}_{4,4}(g) =
\frac{g^5+\frac{9 g^4}{4}+\frac{261 g^3}{80}+\frac{261 g^2}{80}+\frac{9
   g}{4}+1}{g^4+\frac{47 g^3}{20}+\frac{1201 g^2}{400}+\frac{47 g}{20}+1},
\quad
F^{(-1/3)}_{1,1}(g) = 
\frac{1}{\sqrt[3]{\frac{1}{g^3-\frac{3 g^2}{10}-\frac{3 g}{10}+1}}},
\quad
F^{(-1/3)}_{2,2}(g) =
\frac{1}{\sqrt[3]{\frac{g+1}{g^4+\frac{7 g^3}{10}+\frac{243 g^2}{200}+\frac{7
   g}{10}+1}}}
,\NN\\
&&\scriptscriptstyle   
F^{(-1/3)}_{3,3}(g) = 
\frac{1}{\sqrt[3]{\frac{g^2+\frac{49 g}{30}+1}{g^5+\frac{4 g^4}{3}+\frac{81
   g^3}{40}+\frac{81 g^2}{40}+\frac{4 g}{3}+1}}}
,\quad
F^{(-1/3)}_{4,4}(g) =
\frac{1}{\sqrt[3]{\frac{g^3+\frac{87 g^2}{40}+\frac{87 g}{40}+1}{g^6+\frac{15
   g^5}{8}+\frac{243 g^4}{80}+\frac{5589 g^3}{1600}+\frac{243
   g^2}{80}+\frac{15 g}{8}+1}}}
 ,\NN\\
&&\scriptscriptstyle   
F^{(-1/5)}_{2,2}(g)
= 
\frac{1}{\sqrt[5]{\frac{1}{g^5-\frac{g^4}{2}+\frac{103 g^3}{40}+\frac{103
   g^2}{40}-\frac{g}{2}+1}}}, \quad
F^{(-1/5)}_{3,3}(g)
= \frac{1}{\sqrt[5]{\frac{g+1}{g^6+\frac{g^5}{2}+\frac{83 g^4}{40}+\frac{729
   g^3}{400}+\frac{83 g^2}{40}+\frac{g}{2}+1}}},\NN\\
&&\scriptscriptstyle   
F^{(-1/5)}_{4,4}(g)
= \frac{1}{\sqrt[5]{\frac{g^2+\frac{69 g}{40}+1}{g^7+\frac{49
   g^6}{40}+\frac{217 g^5}{80}+\frac{5103 g^4}{1600}+\frac{5103
   g^3}{1600}+\frac{217 g^2}{80}+\frac{49 g}{40}+1}}},\NN\\
&&\scriptscriptstyle   
F^{(-1/7)}_{3,3}(g)
= \frac{1}{\sqrt[7]{\frac{1}{g^7-\frac{7 g^6}{10}+\frac{147 g^5}{40}-\frac{707
   g^4}{400}-\frac{707 g^3}{400}+\frac{147 g^2}{40}-\frac{7 g}{10}+1}}},
\NN\\
&&\scriptscriptstyle   
F^{(-1/7)}_{4,4}(g)
=\frac{1}{\sqrt[7]{\frac{g+1}{g^8+\frac{3 g^7}{10}+\frac{119
   g^6}{40}+\frac{763 g^5}{400}+\frac{45927 g^4}{16000}+\frac{763
   g^3}{400}+\frac{119 g^2}{40}+\frac{3 g}{10}+1}}}, \NN\\
&&\scriptscriptstyle   
F^{(-1/9)}_{4,4}(g)
= 
\frac{1}{\sqrt[9]{\frac{1}{g^9-\frac{9 g^8}{10}+\frac{963
   g^7}{200}-\frac{1281 g^6}{400}+\frac{138663 g^5}{16000}+\frac{138663
   g^4}{16000}-\frac{1281 g^3}{400}+\frac{963 g^2}{200}-\frac{9 g}{10}+1}}}.
\NN\\
\label{F11-basis-list}
\end{\eqa}

\subsection{$F(g) = e^{\frac{1}{g^4}}e^{ g^4}
K_{\frac{1}{4}}(\frac{1}{g^4})
K_{\frac{1}{4}}(g^4)$}
\label{Eq:Ex-Both}
\begin{\eqa}
&&\scriptscriptstyle 
F^{(1)}_{2,2}(g)
= \frac{2.7019 g \left(g^2+1\right)}{g^4+1.95598 g^2+1}
 ,\quad
F^{(1/3)}_{2,2}(g) = 
2.7019 g \left(\frac{1}{g^6+2.86793
   g^4+2.86793 g^2+1}\right)^{\frac{1}{3}}
\NN\\
&&\scriptscriptstyle 
F^{(1/4)}_{2,4}(g) =
2.7019 g \left(\frac{1}{g^8+3.82391
   g^6+5.51393 g^4+3.82391 g^2+1}\right)^{\frac{1}{4}},
\NN\\
&&\scriptscriptstyle 
F^{(1/5)}_{2,6}(g)= 2.7019 g \left(\frac{1}{g^{10}+4.77989
   g^8+9.17715 g^6+8.91928 g^4+4.77989
   g^2+1}\right)^{\frac{1}{5}}
\NN\\
&&\scriptscriptstyle 
F^{(1)}_{4,4}(g) =
\frac{2.7019 g \left(14.4375 g^4+16.9441
   g^2+14.4375\right)}{14.4375 g^6+30.746
   g^4+30.746 g^2+14.4375},
\quad
%\NN\\
%&&\scriptscriptstyle 
F^{(1/3)}_{4,4}(g) = 2.7019 g \left(%\sqrt[3]{
\frac{g^2+1}{g^8+3.86793 g^6+5.63254 g^4+3.86793g^2+1}
\right)^{\frac{1}{3}}
,\NN\\
&&\scriptscriptstyle 
F^{(1/5)}_{4,4}(g) =
2.7019 g 
\left(
\frac{1}{g^{10}+4.77989
   g^8+9.17715 g^6+9.17715 g^4+4.77989
   g^2+1}\right)^{\frac{1}{5}},\NN\\
&&\scriptscriptstyle   
F^{(1/2)}_{4,6}(g) = 2.7019 g \sqrt{\frac{g^4+0.412688
   g^2+1}{g^8+2.32464 g^6+2.71822
   g^4+2.32464 g^2+1}},\NN\\
&&\scriptscriptstyle   
F^{(1/6)}_{4,6}(g) =
2.7019 g %\sqrt[6]{
\left(\frac{1}{g^{12}+5.73587
   g^{10}+13.7543 g^8+17.7363 g^6+13.7543
   g^4+5.73587 g^2+1}
\right)^{\frac{1}{6}} ,\NN\\
&&\scriptscriptstyle   
F^{(1/7)}_{4,8}(g)
= 2.7019 g %\sqrt[7]{
\left(
\frac{1}{g^{14}+6.69184
   g^{12}+19.2453 g^{10}+30.9362
   g^8+30.522 g^6+19.2453 g^4+6.69184
   g^2+1}\right)^{\frac{1}{7}},\NN\\
&&\scriptscriptstyle  
F^{(1/5)}_{6,6}(g)
= 2.7019 g 
\left(\frac{g^2+1}{
   g^{12}+5.77989 g^{10}+13.957
   g^8+18.0964 g^6+13.957 g^4+5.77989
   g^2+1}\right)^{\frac{1}{5}},\NN\\
&&\scriptscriptstyle   
F^{(1/7)}_{6,6}(g)
= 2.7019 g 
\left(\frac{1}{g^{14}+6.69184
   g^{12}+19.2453 g^{10}+30.9362
   g^8+30.9362 g^6+19.2453 g^4+6.69184
   g^2+1}\right)^{\frac{1}{7}},\NN\\
&&\scriptscriptstyle   
F^{(1/9)}_{8,8}(g)
= 2.7019 g 
\left(
\frac{1}{g^{18}+8.6038
   g^{16}+32.9689 g^{14}+73.9794
   g^{12}+107.837 g^{10}+107.837
   g^8+73.9794 g^6+32.9689 g^4+8.6038
   g^2+1}\right)^{\frac{1}{9}}\NN\\
\label{List-Int-Asym}
\end{\eqa}

\providecommand{\href}[2]{#2}\begingroup\raggedright\endgroup


\begin{thebibliography}{10}
\bibitem{Honda:2014bza} 
  M.~Honda,
  %``On Perturbation theory improved by Strong coupling expansion,''
  arXiv:1408.2960 [hep-th].
  %%CITATION = ARXIV:1408.2960;%%


\bibitem{Asnin:2007rw}
V.~Asnin, D.~Gorbonos, S.~Hadar, B.~Kol, M.~Levi, {\em et.~al.}, {\it {High and
  Low Dimensions in The Black Hole Negative Mode}},  {\em Class.Quant.Grav.}
  {\bf 24} (2007) 5527--5540, [\href{http://xxx.lanl.gov/abs/0706.1555}{{\tt
  arXiv:0706.1555}}].

\bibitem{Banks:2013nga}
T.~Banks and T.~Torres, {\it {Two Point Pade Approximants and Duality}},
  \href{http://xxx.lanl.gov/abs/1307.3689}{{\tt arXiv:1307.3689}}.

\bibitem{Kleinert:2001ax}
H.~Kleinert and V.~Schulte-Frohlinde, {\it {Critical properties of
  phi**4-theories}},  {\em River Edge, USA: World Scientific} (2001).


\bibitem{Sen:2013oza}
A.~Sen, {\it {S-duality Improved Superstring Perturbation Theory}},  {\em JHEP}
  {\bf 1311} (2013) 029, [\href{http://xxx.lanl.gov/abs/1304.0458}{{\tt
  arXiv:1304.0458}}].

%\cite{Honda:2014bza}


\bibitem{Pius:2013tla}
R.~Pius and A.~Sen, {\it {S-duality improved perturbation theory in
  compactified type I/heterotic string theory}},  {\em JHEP} {\bf 1406} (2014)
  068, [\href{http://xxx.lanl.gov/abs/1310.4593}{{\tt arXiv:1310.4593}}].



\bibitem{Beem:2013hha}
C.~Beem, L.~Rastelli, A.~Sen, and B.~C. van Rees, {\it {Resummation and
  S-duality in N=4 SYM}},  \href{http://xxx.lanl.gov/abs/1306.3228}{{\tt
  arXiv:1306.3228}}.

\bibitem{Alday:2013bha}
L.~F. Alday and A.~Bissi, {\it {Modular interpolating functions for N=4 SYM}},
  \href{http://xxx.lanl.gov/abs/1311.3215}{{\tt arXiv:1311.3215}}.

%9
\bibitem{Balian:1974xw}
R.~Balian, J.~Drouffe, and C.~Itzykson, {\it {Gauge Fields on a Lattice. 3.
  Strong Coupling Expansions and Transition Points}},  {\em Phys.Rev.} {\bf
  D11} (1975) 2104.

\bibitem{Bali:2014fea}
G.~S. Bali, C.~Bauer, and A.~Pineda, {\it {Perturbative expansion of the
  plaquette to ${\cal O}(\alpha^{35})$ in four-dimensional SU(3) gauge
  theory}},  {\em Phys.Rev.} {\bf D89} (2014) 054505,
  [\href{http://xxx.lanl.gov/abs/1401.7999}{{\tt arXiv:1401.7999}}].

%\cite{DiRenzo:1995qc}
\bibitem{DiRenzo:1995qc} 
  F.~Di Renzo, E.~Onofri and G.~Marchesini,
  ``Renormalons from eight loop expansion of the gluon condensate in lattice gauge theory,''
  Nucl.\ Phys.\ B {\bf 457}, 202 (1995)
  [hep-th/9502095].
  %%CITATION = HEP-TH/9502095;%%

\bibitem{DiRenzo:2000ua}
F.~Di~Renzo and L.~Scorzato, {\it {A Consistency check for renormalons in
  lattice gauge theory: beta**(-10) contributions to the SU(3) plaquette}},
  {\em JHEP} {\bf 0110} (2001) 038,
  [\href{http://xxx.lanl.gov/abs/hep-lat/0011067}{{\tt hep-lat/0011067}}].

%\cite{Horsley:2012ra}
\bibitem{Horsley:2012ra} 
  R.~Horsley, G.~Hotzel, E.~M.~Ilgenfritz, R.~Millo, H.~Perlt, P.~E.~L.~Rakow, Y.~Nakamura and G.~Schierholz {\it et al.},
  ``Wilson loops to 20th order numerical stochastic perturbation theory,''
  Phys.\ Rev.\ D {\bf 86}, 054502 (2012)
  [arXiv:1205.1659 [hep-lat]].
  %%CITATION = ARXIV:1205.1659;%%

%\cite{Di Renzo:2004ge}
\bibitem{Di Renzo:2004ge} 
  F.~Di Renzo and L.~Scorzato,
  ``Numerical stochastic perturbation theory for full QCD,''
  JHEP {\bf 0410}, 073 (2004)
  [hep-lat/0410010].
  %%CITATION = HEP-LAT/0410010;%%

%\cite{Osterwalder:1977pc}
\bibitem{Osterwalder:1977pc} 
  K.~Osterwalder and E.~Seiler,
  %``Gauge Field Theories on the Lattice,''
  Annals Phys.\  {\bf 110}, 440 (1978).
  %%CITATION = APNYA,110,440;%%
  %489 citations counted in INSPIRE as of 09 Aug 2014

\bibitem{Kastening}
B.~Kastening, {\it {Simplified Transfer Matrix Approach in the Two-Dimensional
  Ising Model with Various Boundary Conditions }},  {\em Phys.Rev.} {\bf E66}
  (2002) 057103, [\href{http://xxx.lanl.gov/abs/cond-mat/0209544}{{\tt
  cond-mat/0209544}}].

% 15
\bibitem{Marino:2012zq}
M.~Marino, {\it {Lectures on non-perturbative effects in large N gauge
  theories, matrix models and strings}},
  \href{http://xxx.lanl.gov/abs/1206.6272}{{\tt arXiv:1206.6272}}.


\end{thebibliography}
\end{document}